\documentclass[aps,prb,floatfix,twocolumn,showpacs]{revtex4}
\usepackage{epsfig}
\usepackage{amsmath}
\usepackage{amssymb}
\usepackage{amsfonts}
\usepackage{ dsfont }

\newcommand{\be}{\begin{equation}}
\newcommand{\ee}{\end{equation}}
\newcommand{\vecs}[1]{ \langle {\bf s}({\bf R}_{#1}) \rangle}

\begin{document}


\hsize\textwidth\columnwidth\hsize\csname@twocolumnfalse\endcsname

\bibliographystyle{plain}

\title{Spin-orbit coupled particle in a spin bath}

\author{Peter Stano$^{1,2}$, Jaroslav Fabian$^3$, Igor \v{Z}uti\'{c}$^4$}
\affiliation{
$^1$Institute of Physics, Slovak Academy of Sciences, 845 11 Bratislava, Slovakia\\
$^2$Department of Physics, Klingerbergstrasse 82, University of Basel, Switzerland\\
$^3$Institute for Theoretical Physics, University of Regensburg, D-93040 Regensburg, Germany\\
$^4$Department of Physics, University at Buffalo, NY 14260-1500
}

\vskip1.5truecm
\begin{abstract}

We consider a spin-orbit coupled particle confined in a quantum dot in a bath of impurity spins. We investigate the consequences of spin-orbit coupling on the interactions that the particle mediates in the spin bath. We show that in the presence of spin-orbit coupling, the impurity-impurity interactions are no longer spin-conserving. We quantify the degree of this symmetry breaking and show how it relates to the spin-orbit coupling strength. We identify several ways how the impurity ensemble can in this way relax its spin by coupling to phonons. A typical resulting relaxation rate for a self-assembled Mn-doped ZnTe quantum dot populated by a hole is 1 $\mu$s. We also show that decoherence arising from nuclear spins in lateral quantum dots is still removable by a spin echo protocol, even if the confined electron is spin-orbit coupled.

\end{abstract}
\maketitle

\section{Introduction}

A singly occupied quantum dot is the state of the art of a controllable quantum system in a semiconductor.\cite{loss1998:PRA,hanson2007:RMP}
Coherent manipulation of the particle spin has been demonstrated in lateral dots, where top gates allow for astonishing degree of control by electric fields\cite{taylor2007:PRB, koppens2006:N, nowack2007:S, obata2010:PRB} and in self-assembled dots, where a weaker control over the dot shape and position is compensated by the speed of the optical manipulation.\cite{press2008:N} In both of these major groups, there are two main spin-dependent interactions of the confined particle and the semiconductor environment: spin-orbit coupling embedded in the band structure, and spin impurities, which are either nuclear spins, or magnetic atoms.\cite{fabian2007:APS,zutic2004:RMP} 

A particle couples to an impurity spin dominantly through an exchange interaction, which conserves the total spin of the pair.\cite{coish2009:PSS} This way the electron spin in a lateral GaAs dot will decohere within 10 ns due to the presence of nuclei.\cite{erlingsson2001:PRB, khaetskii2002:PRL, merkulov2002:PRB, coish2006:PSS, cywinski2009:PRB, cywinski2009:PRL, merkulov2010:PRB} Typically, such decoherence is considered a nuisance that can be partially removed by spin echo techniques prolonging the coherence to hundreds microseconds.\cite{petta2005:S, koppens2008:PRL, bluhm2010:N} Whether that decoherence  time can be extended further, e.g. by polarizing the impurities,\cite{schliemann2003:JPCM, reilly2008:S} is not clear, as the experimentally achieved degree of polarization has been so far insufficient, despite great efforts.\cite{bracker2005:PRL, lai2006:PRL, hogele2011:CM} On the other hand, a strong particle-impurity interaction is desired in magnetically doped quantum dots.\cite{seufert2001:PRL, besombes2004:PRL, xiu2010:ACSN, klopotowski2011:PRB, maksimov2000:PRB, beaulac2009:S, sellers2010:PRB, govorov2005:PRB, govorov2008:CRP, fernandez-rossier2004:PRL, qu2006:PRL, nguyen2008:PRB, abolfath2007:PRL, oswaldowski2011:PRL, lebedeva2010:PRB} Here the confined particle is central for both supporting energetically, and assisting in creation, the desired magnetic order of the impurity ensemble. In fact, similar magnetic ordering can be traced to the studies of magnetic polarons in bulk semiconductors, for over fifty years.\cite{yakovlev} The formation of  a magnetic polaron can be viewed as a “cloud” of localized impurity spins, aligned through exchange interaction with a confined carrier spin.\cite{dietl1983:PRB, wolff, nagaev, furdyna1988:JAP}

The conservation of the spin by the impurity-particle interaction is a crucial property. For example, the spin relaxation of the impurity ensemble is impossible with only spin-conserving interactions at hand. This motivates us to consider possibilities to break this symmetry. The first and obvious candidate is the spin-orbit coupling (SOC).\cite{fabian2007:APS,zutic2004:RMP} Despite being weak on the scale of the particle orbital energies, it dominates the relaxation of the particle spin in electronic dots, as is well known,\cite{amasha2008:PRL} because it is the dominant spin-non-conserving interaction. The questions we pose and answer in this work are: assuming the particle is weakly spin-orbit coupled, how strong are the effective spin-non-conserving interactions which appear in the impurity ensemble and what is their form? Is the induced particle decoherence still removable by spin echo? Is the particle efficient in inducing impurity ensemble spin relaxation, thereby limiting the achievable degree of the dynamical nuclear spin polarization\cite{paget1977:PRB}? Can the magnetic order be created through the spin-non-conserving particle mediated interactions --- that is, is this a relevant magnetic polaron creation channel? 

To address these questions we develop here a framework allowing us to treat different particles and impurity spins in a unified manner. We apply our method to two specific systems: a lateral quantum dot in GaAs occupied by a conduction electron with nuclear spins of constituent atoms as the spin impurities, and a self-assembled ZnTe quantum dot occupied by a heavy hole doped with Mn atoms as the spin impurities (readily incorporated as Mn is isovalent with group-II atom Zn). Both of these systems are quasi-two dimensional, the particle spin-orbit coupling is weak compared to the particle orbital level spacing and the particle-impurity interaction is weak compared to the particle orbital and spin level spacings.\cite{hanson2007:RMP, sellers2010:PRB, kuo2006:APL, footnote7} As it is known,\cite{desousa2009:TAP} in this regime one can derive an effective Hamiltonian for the impurity ensemble only, in which the particle does not appear explicitly. This can be done including the particle-bath interaction perturbatively in the lowest order, see the scheme in Fig.~1. Our contribution is in showing how the procedure generalizes to a spin-orbit coupled particle. In addition, we use the resulting Hamiltonian for the calculation of the spin relaxation of the impurities which is phonon-assisted (required to dissipate energy) and particle-mediated (required to dissipate spin). We come up with (and evaluate the corresponding rates for) five possible mechanisms how the spin flips can proceed: shifts of the particle by the phonon electric field (Sec.~IV.B), position shifts of the impurity atoms (Sec.~IV.B), relative shifts of bulk bands (Sec.~IV.C), renormalization of the spin-orbit interactions due to band shifts (Sec.~IV.D) and spin-orbit interactions arising from the phonon electric field (Sec.~IV.E).

Our main findings are the following: 1.~The spin-non-conserving interactions couplings are proportional to the spin-conserving ones multiplied by some power of small parameter(s) which quantify the spin-orbit interaction. For the electron, the small parameter is the dot dimension divided by the spin orbit length and the proportionality is linear. For the hole, the small parameters are the amplitudes of the light hole admixtures into the heavy hole states. The proportionality differs (from linear to quadratic) depending on which hole excited state mediates the interaction. The interaction form is given in Eq.~\eqref{eq:effective hamiltonian final}, our main result. 2.~For the electron, the additional decoherence is removed by the spin echo, while for holes only a partial removal is possible. The latter is because, unlike for electrons, the spin-non-conserving coupling is mediated rather efficiently through higher excited states. 3.~The piezoelectric acoustic phonons are most efficient in relaxing the impurity spin. The resulting relaxation time is unobservably long for nuclear spins, while the hole-induced Mn spin relaxation time of 1 $\mu$s is typical for a 10 nm self-assembled quantum dot, where experimentally measured  times for the polaron creation range from nano to picoseconds.\cite{seufert2001:PRL, beaulac2009:S, sellers2010:PRB} From this we conclude that the interplay of spin-orbit coupling and phonons does not govern the dynamics of magnetic polaron formation at moderate Mn densities (few percents), but rather represents the spin-lattice relaxation timescale, similarly as is the case in quantum wells.\cite{dietl1995:JMMM,yakovlev} Apart from a very low Mn doping, the analytical formulas presented in this work allow us to identify additional regimes where the particle mediated spin relaxation could be relevant for the polaron creation: an example is a hole located at a charged impurity.

The article is organized as follows: In Sec.~II we introduce the description of the particle focusing on the spin-orbit coupling. In Sec.~III we specify the particle-impurity interaction, define its important characteristics and derive the effective Hamiltonian for the impurity ensemble. In Sec.~IV we calculate the spin relaxation rates for the impurity ensemble, after which we conclude. We put numerous technical details into Appendices, with which the text is self-contained.

\section{Quantum dot states and spin-orbit interaction}

\begin{figure}
\includegraphics[width=0.45\textwidth]{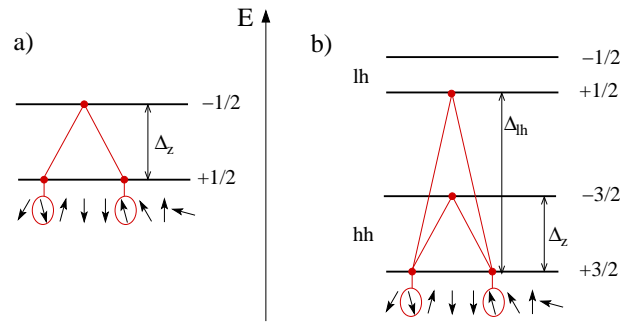}
\caption{(Color online) Effective interaction between impurities (encircled in red/gray) mediated by a confined particle (red/gray lines). a) Electron is excited from the ground state of spin +1/2 into the closest spin -1/2 state (up by the Zeeman energy $\Delta_z$) upon flipping one of the impurities and deexcited back upon flipping another one. b) Hole spectrum is more complicated, comprising heavy hole (hh) and light hole (lh)-like states, the latter displaced by light-heavy hole splitting $\Delta_{lh}$.  
\label{fig:scheme} }
\end{figure}

\subsection{Electron states}

In the single band effective mass approximation, that we adopt, the Hamiltonian of a quantum dot electron is
\begin{equation}
H_{\rm dot} = \frac{{\bf p}^2}{2m}  + V ({\bf r}) + \frac{p_z^2}{2m}  + V_z (z) + g\mu_B {\bf B} \cdot {\bf J} +H_{\rm so}.
\label{eq:electronic quantum dot hamiltonian}
\end{equation}
The underlying bandstructure is taken into account as a renormalization of the mass $m$ and the $g$ factor in the electron kinetic and Zeeman energies, respectively. The latter couples the external magnetic field ${\bf B}$ to the electron spin through the vector of Pauli matrices $\boldsymbol{\sigma}=2{\bf J}$. We will below assume a sizeable (above 100 mT) external magnetic field in the electronic case, which is typical in experiments to allow for the electron spin measurement\cite{elzerman2004:N, amasha2008:PRL} and to slow down the impurity dynamics and the resulting decoherence.\cite{cywinski2009:PRL} On the other hand, we neglect the orbital effects of this field which is justified if the confinement length is much smaller than the magnetic length $l_B=\surd{2\hbar/eB}$, where $-e$ is the electron charge. If the field is strong (above 1-2  Tesla), the orbital effects important here are fully incorporated by a renormalization of the confinement length $l^{-4} \to l^{-4}+l_B^{-4}$. The field is applied along ${\bf s}_0$, a unit vector.

The quantum dot is defined by the confinement potential $V+V_z$, which we separated into the in-plane and perpendicular components. The corresponding in-plane and perpendicular position and momentum components read ${\bf R}=\{{\bf r},z\}$ and ${\bf P}=\{{\bf p},p_z\}$, respectively. Whenever we below need an explicit form of the wavefunction, we assume, for convenience, a parabolic in-plane and a hard-wall perpendicular confinements
\begin{subequations}
\label{eq:potential}
\begin{eqnarray}
V({\bf r}) &=& \frac{\hbar^2}{2 m l^4} r^2 \equiv \frac{1}{2} m \omega^2 r^2,\\
V_z(z) &=& \left\{ 
\begin{tabular}{ll}
0,&$0<z<w$,\\
$\infty$,& otherwise.
\end{tabular}
\right.
\end{eqnarray}
\end{subequations}
The confinement lengths $l$ and $w$ characterize a typical extent of the wavefunction in the lateral, and perpendicular directions, respectively. Alternatively to the length, the confinement energy $\hbar \omega$ can be used. 
However, we stress that our results below do not rely on the specific confinement form in any way, as long as the dot is quasi two dimensional, a condition which for the adopted example reads $l\gg w$. Typical values for lateral quantum dots in GaAs are $l$=30 nm, $w$=8 nm.

The last term in Eq.~\eqref{eq:electronic quantum dot hamiltonian} is the spin-orbit interaction\cite{fabian2007:APS}
\begin{equation}
H_{\rm so} = \frac{\hbar}{2 m l_{\rm d}}(-\sigma_x p_x + \sigma_y p_y)+\frac{\hbar}{2 m l_{\rm br}}(\sigma_x p_y - \sigma_y p_x),
\label{eq:electronic spin orbit}
\end{equation}
comprising the Dresselhaus term, which arises in zinc-blende structures grown along [001] axis and the Bychkov-Rashba term, which is a consequence of the strong perpendicular confinement. The interactions are parameterized by the spin-orbit lengths $l_{\rm so}\in\{l_{\rm d},l_{\rm br}\}$, typically a few microns in GaAs heterostructures. 

Assume first that the spin-orbit coupling is absent. To be able to treat the electron and the hole (each referred to as the particle) on the same footing, we introduce the following notation
\begin{equation}
|\Psi_p\rangle = |J\rangle \otimes |\Phi^J_a\rangle, \qquad \textrm{(zero SOC)}.
\label{eq:electronic basis 0}
\end{equation}
The complete particle wavefunction, for which we use the Greek letter $\Psi$, is a two (electron) or four component (hole) spatially dependent spinor. Its label $p=\{J, a\}$ indicates that the wavefunction is separable into a (position independent) spinor $|J\rangle$ and a scalar position dependent complex amplitude $|\Phi\rangle$. The former is labelled by the particle angular momentum in units of $\hbar$, $J=\pm$1/2 (electron), $J=\pm$3/2,$\pm$1/2 (hole; alternatively, we use $hh$ for 3/2 and $lh$ for 1/2 labels). The set of orbital quantum numbers $a$ depends on the confinement potential. For the choice in Eqs.~\eqref{eq:potential}, it is a set of three numbers $a=\{nm,k\}$, with $n$ the main and $m$ the orbital quantum number ($\overline{m}\equiv-m$) of a Fock-Darwin state, and $k$ the label of the subband in the perpendicular hard-wall confinement. Finally, for the particle ground state we omit the index $a$, or  use $G\equiv \{J,00,0\}$ in place of $p$. The electron ground state is thus denoted by
\be
|\Psi_{1/2}\rangle = |1/2\rangle \otimes |\Phi_G\rangle,
\ee
where the direction of the spin up spinor $|1/2\rangle$ is set by the external field along ${\bf s}_0$.

Let us now consider the spin-orbit coupling. It turns out that for electrons the spin-orbit effects on the wavefunction can be in the leading order written as\cite{aleiner2001:PRL,levitov2003:PRB}
\begin{equation}
|\Psi_p\rangle = U \,|J\rangle \otimes |\Phi^J_a\rangle.
\label{eq:electronic basis 1}
\end{equation}
Here $U$ is a unitary 2 by 2 matrix of a spinor rotation, 
\begin{equation}
U({\bf r})=\exp\left[-{\rm i} \, {\bf n}_{\rm so}({\bf r})\cdot {\bf J} \right],
\label{eq:transformation U}
\end{equation}
parameterized by an in-plane position dependent vector
\begin{equation}
{\bf n}_{\rm so} ({\bf r}) = -\left( \frac{x}{l_{\rm d}}-\frac{y}{l_{\rm br}}, \frac{x}{l_{\rm br}}-\frac{y}{l_{\rm d}},0 \right).
\label{eq:effective spin-orbit vector}
\end{equation}
A weak spin-orbit coupling allows us to label states with the same quantum numbers as for no spin-orbit case, as there is a clear one to one correspondence. The enormous simplification, that the unitary matrix in Eq.~\eqref{eq:electronic basis 1} does not depend on the quantum numbers $p$, is due to the special form of the spin-orbit coupling in Eq.~\eqref{eq:electronic spin orbit}, what has several interesting consequences.\cite{stepanenko2003:PRB,schliemann2003:PRL,tokatly2010:AP,tokatly2010:PRB} To calculate the spin relaxation or spin-orbit induced energy shifts, on which $U$ has no effects, one has to go beyond the leading order given in Eq.~\eqref{eq:electronic basis 1}.\cite{stano2006:PRL} However, we will see that here $U$ will result in spin-non-conserving interactions, and it is thus enough to consider the leading order. For the same reason we neglected the cubic Dresselhaus term in Eq.~\eqref{eq:electronic spin orbit}, what is here, unlike usually,\cite{baruffa2010:PRL} an excellent approximation.

\subsection{Hole states}

For holes we restrict to the four dimensional subspace of the light and heavy hole subbands. Neglecting the spin-orbit coupling, they correspond to the angular momentum states $J=\pm3/2$, and $J=\pm1/2$, respectively. We use the confinement potential in Eqs.~\eqref{eq:potential}, setting the confinement energy in the heavy hole subband to 20 meV, which gives $l\approx 4$ nm and setting the light-heavy hole splitting to $\Delta_{lh}=100$ meV, which gives $w\approx 2$ nm.
The atomic spin-orbit coupling manifests itself as the orbital splitting of the light and heavy holes from the spin-orbit split-off band (which is energetically far from the states considered in this paper), and as a coupling of the light and heavy holes at finite momenta. The latter effect we refer to as the hole spin-orbit coupling --- we do not consider higher order effects, which give rise to spin-orbit interactions similar in form to the electronic Dresselhaus and Rashba terms. Within this model, we derive the spin-orbit coupled wavefunctions from the corresponding 4 by 4 sector of the Kohn-Luttinger Hamiltonian in Appendix \ref{app:KLH}, and get the hole ground state as
\begin{equation}
\begin{split}
|\Psi_{3/2}\rangle  =&  |3/2\rangle\otimes|\Phi^{hh}_{00,0}\rangle + \lambda_1 |1/2\rangle \otimes |\Phi^{lh}_{01,1}\rangle \\&+ \lambda_0  |-1/2\rangle \otimes |\Phi^{lh}_{02,0}\rangle.
\end{split}
\label{eq:1-st order ground state}
\end{equation}
We use the notation explained below Eq.~\eqref{eq:electronic basis 0}. The Fock-Darwin states in the heavy hole ($hh$) and light hole ($lh$) subband differ, due to different effective masses. The key quantities are the scalars $\lambda$, which quantify the light-heavy hole mixing. The spin-non-conserving interactions, as well as the resulting spin relaxation rates, will scale with these scalars. For our parameters, which we list for convenience in Appendix \ref{app:parameters}, we get $\lambda_0\approx\lambda_1\approx0.05$. In addition to the ground state, we will need also the lowest excited state in the heavy hole subband, which is the time reversed copy of Eq.~\eqref{eq:1-st order ground state},
\begin{equation}
\begin{split}
|\Psi_{-3/2}\rangle  =&  |-3/2\rangle\otimes|\Phi^{hh}_{00,0}\rangle - \lambda_1 |-1/2\rangle \otimes |\Phi^{lh}_{0\overline{1},1}\rangle \\&+ \lambda_0  |1/2\rangle \otimes |\Phi^{lh}_{0\overline{2},0}\rangle,
\end{split}
\label{eq:1-st order ground state 2}
\end{equation}
and also the lowest state in the light hole subband
\begin{equation}
\begin{split}
|\Psi_{1/2}\rangle  =&  |1/2\rangle\otimes|\Phi^{lh}_{00,0}\rangle + \lambda_1^\prime |3/2\rangle \otimes |\Phi^{hh}_{0\overline{1},1}\rangle \\&- \lambda_0^\prime  |-3/2\rangle \otimes |\Phi^{lh}_{0\overline{2},0}\rangle,
\end{split}
\label{eq:1-st order ground state 3}
\end{equation}
which will be surprisingly effective in inducing the spin-non-conserving coupling among impurities, as we will see. We make a few notes here: First, in the spherical approximation that we adopt, the Kohn-Luttinger Hamiltonian conserves the angular momentum, so that all components in each of Eqs.~\eqref{eq:1-st order ground state}-\eqref{eq:1-st order ground state 3} have the same value of $J+m$. Second, the mixing is stronger in the light hole subband, $\lambda_1^\prime \approx 0.15$ and  $\lambda_0^\prime \approx 0.11$. This is because the admixing states are closer in energy---the Fock-Darwin excitation energies add to and subtract from the light-heavy hole splitting in the heavy and light hole case, respectively, as evidenced by Eqs.~\eqref{eq:lambdas} and \eqref{eq:lambdas prime}. Third, we will be interested in the case of zero external magnetic field for holes (we give the Zeeman interaction in Appendix \ref{app:KLH} for completeness.) Unlike for electrons, such a field is here not required to split the two states in Eqs.~\eqref{eq:1-st order ground state} and \eqref{eq:1-st order ground state 2}, as the splitting arises due to the spin impurities. As we will see, this splitting will be of the order of few meV. Compared to this, the hole Zeeman energy is negligible up to fields of several Tesla. In addition, the external field suppresses an interesting feedback between the particle and impurities.\cite{oswaldowski2012:U} Finally, we note that one could relate the first-order and unperturbed hole states analogously to the electron case introducing a unitary transformation $U$, whose matrix elements are the coefficients appearing in Eqs.~\eqref{eq:1-st order ground state}-\eqref{eq:1-st order ground state 3}. However, since here the transformation does not have any appealing form similar to the one in Eq.~\eqref{eq:transformation U}, we do not explicitly construct the matrix $U$ for holes.

\section{Effective Hamiltonian}

In this section we introduce the particle-impurity exchange interaction, while providing a unified description for both type of particles and impurities. This interaction manifests itself as the Knight field acting on the impurities and the Overhauser field acting on the particle. (The fields are defined as the expectation value of the exchange interaction in the state of the corresponding subsystem). Historically, the terminology was initially applied to nuclei and here we also use it for Mn spins. With the help of these fields, we define the unperturbed  basis of the particle-impurity system, for which we derive the effective interaction Hamiltonian treating the non-diagonal exchange terms perturbatively. Finally, we define the spin-conserving versus spin-non-conserving interaction terms and analyze their relative strength for the electron and hole case.

Our strategy can be viewed also in the following alternative way. To derive the spin-orbit coupling effects on the effective impurity interactions, we proceed in two steps: First we unitarily transform the particle basis to remove the spin-orbit coupling in the lowest order. The spin-orbit coupled basis transformation renormalizes the particle-impurity exchange interaction and breaks its spin-rotational symmetry. Then we integrate out the particle degrees of freedom by a second unitary transformation, using the L\"owdin (equivalently here, the Schrieffer-Wolff) transformation, which leaves us with effective interactions concerning impurities only.

\subsection{Particle-impurity interaction}

The particle interacts with impurities by the Fermi contact interaction\cite{paget1977:PRB,fischer2008:PRB}
\begin{equation}
H_F = \sum_n H_F^n = - \sum_n \beta \, \delta({\bf R} - {\bf R}_n) \, {\bf J} \cdot {\bf I}^n.
\label{eq:Fermi contact interaction}
\end{equation}
Here $n=1, 2 \ldots$ labels the impurities located at positions ${\bf R}_n=({\bf r}_n,z_n)$ with corresponding spin operators ${\bf I}^n$ in units of $\hbar$. The impurities have spin $I$ and density $1/v_0$. For the electron-nuclear spins case the impurity density is in GaAs the atomic density. For holes-Mn spins we assume $x_{\rm Mn}$ is the fraction of cations replaced by Mn atoms, typically $x_{\rm Mn}=1\%$. For impurities with different magnetic moments (as for nuclei of different elements) the coupling $\beta$ should have the index $n$, but we will not consider this minor complication. Even though the wavefunction extends to infinity, it makes sense to consider the number of impurities with which the particle interacts appreciably as $N=V/v_0$, where the dot volume $V$ is defined by\cite{merkulov2002:PRB}
\be 
1/V = \int {\rm d}^3{\bf R} |\Phi_G({\bf R})|^4.
\ee
The maximal value of the Hamiltonian $H_F$, if all impurities are aligned with the particle, is
\begin{equation}
E = - \sum_n \beta \, |\Phi_G({\bf R}_n)|^2 J I = -\beta J I / v_0.
\end{equation}
The impurity Zeeman energy is 
\begin{equation}
H_{nZ}= \sum_n g_n \mu_{\rm imp}  {\bf I}^n \cdot {\bf B},
\end{equation}
with $g_n$ the impurity g-factor. For GaAs lateral dots, we consider nuclear spins as impurities, with $I=3/2$ and $\mu_{\rm imp}$ the nuclear magneton $\mu_N$. For ZnTe dots, the impurities are intentionally doped Mn atoms, with $I=5/2$ and $\mu_{\rm imp}$ the Bohr magneton $\mu_B$.

\subsection{Knight field}

Assume the particle sits in the ground state $G$. In the lowest order of the the particle-impurity interaction, a particular impurity spin couples to a local field, called the Knight field.  We define it in units of energy by writing
\begin{equation}
{\bf K}^n \cdot {\bf I}^n \equiv \langle \Psi_G | H_F^n | \Psi_G \rangle,
\label{eq:Knight field definition}
\end{equation}
from which, using Eq.~\eqref{eq:Fermi contact interaction}, we get
\begin{equation}
{\bf K}^n= -\beta \langle \Psi_G | \delta({\bf R}-{\bf R}_n)\, {\bf J}\, | \Psi_G \rangle.
\label{eq:Knight field}
\end{equation}
Using Eq.~\eqref{eq:electronic basis 1} the Knight field of an electron is 
\begin{equation}
\begin{split}
{\bf K}^n &= -\beta | \Phi_G({\bf R}_n) |^2 \langle 1/2 | U^\dagger({\bf R}_n)\, {\bf J}\, U({\bf R}_n) | 1/2 \rangle \\
& = -\beta(1/2) | \Phi_G({\bf R}_n) |^2 \vecs{n}.
\end{split}
\label{eq:Knight field electrons}
\end{equation}
It points along the direction of the electron spin at the position of the $n$-th impurity, introduced as a unit vector $\vecs{n} \equiv R_{U({\bf R}_n)} [{\bf s}_0]$. The operator $R_U$ is defined such that it performs the same rotation on vectors, as $U$ does on spinors. The explicit form of $R$ is the one in Eq.~\eqref{eq:transformation U}, if generators of rotations in three dimensions are used, $(J_k)_{lm} = -{\rm i}\, \epsilon_{klm}$. As is apparent from Eq.~\eqref{eq:Knight field electrons}, evaluating the Knight field with perturbed electron wavefunctions is equivalent to evaluating a unitarily transformed interaction $H_F\to U^\dagger H_F U$ with unperturbed electron wavefunctions.

We get the Knight field of the hole as (see Appendix \ref{app:hole matrix elements}),
\begin{equation}
[ K^n_x, K^n_y, K^n_z] = -\beta [ \lambda_1 {\rm Re}\,f_n, \, \lambda_1 {\rm Im}\, f_n, 3|\Phi_{00,0}^{hh}({\bf R}_n)|^2/2 ] ,
\label{eq:Knight field holes}
\end{equation}
where we abbreviated $f_n=\sqrt{3}\, \Phi_{01,1}^{lh}({\bf R}_n) \Phi_{00,0}^{hh*}({\bf R}_n)$ and neglected contributions of higher orders in $\lambda$.  Rather than the exact form, we note that without the spin-orbit coupling, the direction of the Knight field is fixed globally (along the external magnetic field for electrons, ${\bf s}_0={\bf B}/B$ and along the z axis --- the spin direction of heavy holes --- for holes, ${\bf s}_0 = {\bf \hat{z}}$). The spin-orbit interaction deflects the Knight field in a position dependent way, the deflection being in the leading order linear in the small parameter characterizing the spin-orbit interaction. In this respect, Eq.~\eqref{eq:Knight field electrons} and \eqref{eq:Knight field holes} are the same.

\subsection{Basis}

The total field aligning the impurity spin is the sum of the Knight field and the external field
\be
\boldsymbol{\mathcal{B}}^n = {\bf K}^n + g_n \mu_{\rm imp} {\bf B}.
\label{eq:total effective field}
\ee
The typical energy scale of the Knight field of an electron in a lateral dot is tens of peV, which corresponds to the impurity in an external field of 1 mT. For a hole in a self-assembled dot the Knight field is of the order of 100 $\mu$eV, corresponding to the external field of 0.3 T. Based on this, in the following analysis we mostly consider typical situations, in which the total field is dominated by the external field for nuclear spins (electronic case), and the Knight field for Mn spins (hole case).

We now introduce for each impurity a rotated (primed) coordinate system, in which the unit vector ${\bf \hat{z}^\prime}$ is along the total field $\boldsymbol{\mathcal{B}}^n$. Formally, the rotation is performed by operator $R_{\boldsymbol{\mathcal{B}}^n}$ defined by the relation between the unit vectors,
\be
{\bf \hat{z}^\prime} = R_{\boldsymbol{\mathcal{B}}^n} [ {\bf \hat{s}}_0 ].
\label{eq:local coordinates}
\ee
The orientation of the in-plane axes $x^\prime$, $y^\prime$ in the plane perpendicular to ${\bf \hat{z}^\prime}$ is arbitrary, and we denote ${\bf r^\prime}_\pm = {\bf \hat{x}^\prime} \pm {\rm i} {\bf \hat{y}^\prime}$. We define the impurity ensemble basis states as tensor products of states with a definite spin projection along the locally rotated axis $z^\prime$,
\begin{equation}
|\mathcal{I} \rangle = | I^1_{z^\prime} \rangle \otimes | I^2_{z^\prime} \rangle \otimes \ldots \otimes | I^N_{z^\prime} \rangle.
\label{eq:impurity basis}
\end{equation}
The spin projections take discrete values, $I^n_{z^\prime}\in \{I,I-1,\ldots, -I\}$. We use $\mathcal{I}$ as the collective index of the impurities. With this, we are now ready to introduce the complete system basis, as spanning the states
\begin{equation}
|\Psi_p \rangle \otimes | \mathcal{I} \rangle  \equiv
| \Psi_p \rangle \otimes | I^1_{z^\prime} \rangle \otimes | I^2_{z^\prime} \rangle \otimes \ldots \otimes | I^N_{z^\prime} \rangle,
\end{equation}
with the corresponding energies 
\begin{equation}
E_{p,\mathcal{I}} = E_p + \sum_n E_{I^n_{z^\prime}} = E_p + \sum_n \mathcal{B}^n I^n_{z^\prime},
\end{equation}
comprising the particle energy and the Zeeman energies of impurities in the corresponding total fields.

\subsection{The substantial gap assumption -- Overhauser field}

In addition to the Knight field, another consequence of the particle-impurity interaction from Eq.~\eqref{eq:Fermi contact interaction} is the effective field experienced by the particle spin, known as the Overhauser field\cite{hanson2007:RMP} ${\bf O}$. To express this field again in the units of energy, it is helpful to consider the matrix elements of the particle-impurity interaction within the subspace of the lowest two electron states $J,J^\prime \in S\equiv \{1/2,-1/2\}$,
\be \begin{split}
\langle \Psi_J | &-\beta \sum_n \delta({\bf R}-{\bf R}_n)\, {\bf J} \cdot {\bf I}^n |\Psi_{J^\prime}\rangle \\
&= - \beta \sum_n |\Phi_G({\bf R}_n)|^2 R_{U_n^\dagger}[{\bf I}^n] \cdot \langle J | {\bf J} | J^\prime \rangle.
\end{split}
\label{eq:aux}
\ee
We introduce the field ${\bf O}$ as
\be
H_F|_{S} \equiv {\bf O} \cdot {\bf J},
\ee 
where the subscript $S$ refers to the subspace comprising a pair of time-reversed particle states and the Overhauser field depends on the choice of $S$. To quantify the Overhauser field, we give up on trying to track the microscopic state of the impurities and instead introduce the averaging (denoted by an overline) over  impurity ensembles  
\be
\overline{I_a^n}=0,\qquad \overline{I_a^n I_b^m} = \delta_{nm}\delta_{ab} I(I+1)/3,
\label{eq:impurity ensemble average}
\ee
which characterizes unpolarized and isotropic ensembles. Nuclear spins, unless intentionally polarized in dynamical nuclear polarization schemes,\cite{pfund2007:PRL,reilly2008:S, rudner2007:PRL} usually well fulfill this. If not polarized by an external field, Mn spins can be considered random initially, before the particle enters the dot and the polarization starts to build up.

Equation \eqref{eq:impurity ensemble average} gives a zero Overhauser field on average, but with a finite dispersion, quantifying a typical value. For electrons, we get the well known result,\cite{merkulov2002:PRB, erlingsson2002:PRB}
\be
\begin{split}
\overline{ {\bf O}^2} &= \beta^2 \sum_{nm} |\Phi_G({\bf R}_n)|^2 |\Phi_G({\bf R}_m)|^2 \overline{ R_{U_n^\dagger}[{\bf I}^n] \cdot R_{U_m^\dagger}[{\bf I}^m]} \\
& = \beta^2 \sum_n |\Phi_G({\bf R}_n)|^4 I(I+1) = I(I+1) (\beta/v_0)^2 /N,
\end{split}
\label{eq:Overhauser electron}
\ee
stating that the typical value of the Overhauser field is inversely proportional to the square root of the number of impurity spins within the dot. The spin-orbit coupling, equivalent to a position dependent spin coordinate frame rotation, does not influence the result at all, as Eq.~\eqref{eq:impurity ensemble average} assumes isotropic non-interacting impurities. For our parameters, the typical Overhauser field value is $0.3\,\mu$eV, which corresponds to external field of 20 mT. The energy splitting of the electron spin opposite states is therefore for our case dominated by the Zeeman, rather than the Overhauser, field.

For holes, we will not write the Overhauser field explicitly as a vector. Instead we calculate directly the typical matrix elements of the particle-impurity interaction within the heavy hole subspace with the spin-orbit renormalized wavefunctions. We leave the details for Appendix \ref{app:hole matrix elements} and state the results here: the diagonal terms are
\be
\begin{split}
\overline{|\langle \Psi_{\pm 3/2} | H_F | \Psi_{\pm 3/2} \rangle|^2} & \approx   (3/4) I(I+1) (\beta/v_0)^2 / N,
\end{split}
\ee
where we neglected small contributions of the spin-orbit coupling.
An important difference to an analogous result for the electrons, Eq.~\eqref{eq:Overhauser electron} is the energy scale. Here the typical value for the diagonal Overhauser field is several meV, which corresponds to huge external fields of many Tesla. The energy splitting of the hole is thus dominated by the Overhauser, rather than Zeeman, field. On the other hand, the off-diagonal element is non-zero only in the presence of the spin-orbit coupling, 
\be
\begin{split}
\overline{| \langle \Psi_{-3/2} | H_F | \Psi_{3/2} \rangle|^2} &\sim 2I(I+1) (\lambda_0\beta/v_0)^2 / N.
\end{split}
\label{eq:transversal Overhauser}
\ee
The impurity spins may induce transition (precession) of the heavy hole spin due to the transversal  
component of the Overhauser field, which is smaller by a factor of $\lambda_0$ compared to the diagonal component.  For our parameters the transversal component is of the order of tens of $\mu$eV, so for the heavy hole spin precession to occur, the two spin opposite states have to be degenerate with respect to this energy [which normally does not occur, because of the diagonal term, Eq.~\eqref{eq:Overhauser electron}].

Having compared the typical energy splittings of the particle induced by the effective Overhauser field, versus the external magnetic field, we are now ready to discuss the crucial assumption for the derivation which will follow. It is the assumption that the particle is fixed to its ground state by an energy gap, irrespective of the evolution of the impurity ensemble. This requires that spin flips of impurities cost much less in energy than the particle transitions
\be
\Delta E_{I^n} \ll \Delta E_p.
\label{eq:gap assumption}
\ee
For electrons, this assumption is guaranteed as both the particle and impurities spin flip costs are dominated by the Zeeman energy, proportional to the magnetic moment, which is much larger for the electron than for a nuclear spin, $\mu_{\rm imp}=\mu_N \sim 10^{-3} \mu_B$. On the other hand, for holes for which the particle and impurity magnetic moments are comparable, the above condition is also fulfilled since the particle spin flip energy cost is dominated by the Overhauser field. 

\subsection{The effective Hamiltonian}

Once the particle is fixed to its ground state (the substantial gap assumption), the particle excited states can be integrated out perturbatively\cite{shenvi2005:PRB, yao2006:PRB, desousa2009:TAP} resulting in an effective Hamiltonian for the impurity ensemble $H_{\rm eff}$. For this purpose we split the interaction Hamiltonian to
\be
H_F \equiv H_F^0 + H_F^\prime,
\ee
where the diagonal part,
\be 
H_F^0 = \langle \Psi_G | H_F | \Psi_G \rangle = \sum_n {\bf K}^n \cdot {\bf I}^n,
\ee
together with the external field, defines the unperturbed Hamiltonian $H_0=H_p+H_{nZ} + H^0_F$ and the basis, so that
\be
\langle \Psi_p \otimes \mathcal{I}_A | H_{nZ} + H_F^0 | \Psi_q \otimes \mathcal{I}_B \rangle \propto \delta_{pq} \delta_{AB},
\ee
where $\mathcal{I}_A$, $\mathcal{I}_B$ denote arbitrary basis states of the impurity ensemble. We also note that
\be
\langle \Psi_G \otimes \mathcal{I}_A | H_F^\prime | \Psi_G \otimes \mathcal{I}_B \rangle =0.
\ee
Using L\"owdin theory,\cite{lowdin1951:JCP,bir-pikus} the matrix elements of the effective Hamiltonian, in the lowest order in the non-diagonal part, $H_F^\prime$, are
\begin{widetext}
\begin{equation}
\langle \mathcal{I}_A | H_{\rm eff}| \mathcal{I}_B \rangle = 
\langle \Psi_G \otimes \mathcal{I}_A | \quad H_0 
+\sum_{p\neq G, \mathcal{I}^* }  
\Big(\frac{1/2}{E_{G\mathcal{I}_A} - E_{p\mathcal{I}^*}}+\frac{1/2}{E_{G\mathcal{I}_B} - E_{p\mathcal{I}^*}}\Big) H_F^\prime |\Psi_p\otimes \mathcal{I}^* \rangle \langle \Psi_p \otimes \mathcal{I}^*| H_F^\prime
\quad| \Psi_G \otimes \mathcal{I}_B \rangle,
\end{equation}
\end{widetext}
where the summation proceeds through the excited particle states and a complete basis of impurities. We now use the substantial gap assumption, which states that all system states reachable by the interaction $H_F^\prime$ have the energy dominated by the particle energy, so that we can approximate $E_{G\mathcal{I}}-E_{p\mathcal{I}^*} \approx E_G - E_p$. By this relation the summation over the impurities gives an identity,
\begin{equation}
\begin{split}
&\langle \mathcal{I}_A | H_{\rm eff}| \mathcal{I}_B \rangle =\\& \,\,\,
\langle \Psi_G \otimes \mathcal{I}^A | H_0 
+H_F^\prime 
\sum_{p\neq G} \frac{|\Psi_p \rangle \langle \Psi_p |} {E_G - E_p } 
H_F^\prime
| \Psi_G \otimes \mathcal{I}^B \rangle.
\end{split}
\end{equation}
Since the impurity states now only sandwich both sides of the equation, we can equate the operators
\begin{equation}
H_{\rm eff}=  \langle \Psi_G |H_0| \Psi_G \rangle +\sum_{p\neq G}   
\frac{\langle \Psi_G | H_F^\prime |\Psi_p \rangle \langle \Psi_p | H_F^\prime | \Psi_G \rangle} {E_G - E_p }.
\label{eq:derivation1}
\end{equation}
Even though this looks like the standard second order perturbation theory result, note that even after taking matrix elements with respect to the particle states, the expressions still contain the quantum mechanical operators of the impurity spins. On the other hand, by taking the expectation value, the particle degrees of freedom disappear from the effective Hamiltonian. The first term is a sum of the particle ground state energy and the impurities energy in the Knight field,
\be
\langle \Psi_G |H_0| \Psi_G \rangle = E_G + \sum_n \boldsymbol{\mathcal{B}}^n \cdot {\bf I}^n.
\ee
To simplify the notation of the second term in Eq.~\eqref{eq:derivation1}, we introduce
\be
\langle \Psi_G | H_F^\prime |\Psi_p \rangle = \langle \Psi_G | H_F |\Psi_p \rangle
\equiv \sum_n {\bf A}^n \cdot {\bf I}^n,
\ee
so that the $p$-state dependent complex vector ${\bf A}$  is
\be
{\bf A}^n= -\beta \langle \Psi_G | \delta({\bf R}-{\bf R}_n) \, {\bf J}  | \Psi_p \rangle.
\ee
We now transform vectors ${\bf A}$ and spin operators ${\bf I}$ into the coordinate system along the total field of each impurity
\be
\tilde{\bf A}^n = R_{\boldsymbol{\mathcal{B}}^n}^{-1} [{\bf A}], \qquad
\tilde{\bf I}^n = R_{\boldsymbol{\mathcal{B}}^n}^{-1} [{\bf I}].
\ee
The z-component of a rotated vector is its projection along the direction of the local total field, e.g. $\tilde{I}_z = {\bf I}\cdot {\bf \hat{z}^\prime}$ and similarly for ${\bf A}$. Omitting the constant $E_G$, we rewrite Eq.~\eqref{eq:derivation1} with the new notation and arrive at our main result 
\begin{equation}
H_{\rm eff}=  \sum_n \mathcal{B}^n \tilde{I}^n_z +\sum_{p\neq G} \sum_{n,m}  
\frac{1}{E_G - E_p }
(\tilde{\bf A}^{n}  \cdot \tilde{\bf I}^{n}) (\tilde{\bf A}^{m} \cdot \tilde{\bf I}^{m})^\dagger .
\label{eq:effective hamiltonian final}
\end{equation}
The first term defines the spin flip energy cost and the spin quantization axis, according to Eq.~\eqref{eq:local coordinates}. The interactions described by the second term can be classified as spin-conserving (spin-non-conserving) according to rotated operators $\tilde{\bf I}$ components parallel (perpendicular) to a global axis ${\bf \hat{s}}_0$, as we will show below. To further demonstrate the usefulness and generality of Eq.~\eqref{eq:effective hamiltonian final}, we show that known results follow as special limits, and how the consequences of the spin-orbit coupling on the impurities interactions can be drawn from the formula. We also note that the derivation would proceed in the same way even if $G$ were not the particle ground state. The only requirement for the validity of Eq.~\eqref{eq:effective hamiltonian final} is that the state $G$ is far enough in energy from other particle states so that Eq.~\eqref{eq:gap assumption} is valid. For example, thermal excitations of the particle would result in a thermal average of the effective Hamiltonian (the vectors ${\bf A}$ and energies $\mathcal{B}$ do depend on $G$). We do not pursue a finite temperature regime further here, and assume the thermal energy $k_B T$ is small such that the particle stays in the ground state.

Before we evaluate vectors $\tilde{\bf A}$ in specific cases, we note an important property of the effective Hamiltonian. Namely, for both holes and electrons, the lowest excited state is much closer to the ground state (split by the Zeeman energy) compared to higher excited states (split by orbital excitation energies). If the mediated interactions are dominated by this low lying excited state, we can write
\begin{equation}
H_{\rm eff}=\sum_n \mathcal{B}^n \tilde{I}^n_z + \sum_{n,m}  
\frac{1}{E_\uparrow - E_\downarrow }
(\tilde{\bf A}^{n}_{\uparrow\downarrow}  \cdot \tilde{\bf I}^{n}) (\tilde{\bf A}^{m}_{\uparrow\downarrow} \cdot \tilde{\bf I}^{m})^\dagger,
\label{eq:effective hamiltonian approximation}
\end{equation}
where we denoted $G=\uparrow$, the sum was restricted to a single term $p=\downarrow$, and  $\tilde{\bf A}^{n}_{\uparrow\downarrow}$ is the  vector corresponding to the ground/excited state being the spin up/down state. From the relation $\tilde{\bf A}^{n}_{\uparrow\downarrow} = (\tilde{\bf A}^{n}_{\downarrow\uparrow})^\dagger$ it is straightforward to see that had we started with the particle being in the spin down state $G=\downarrow$ and restricted to the closest state in the spectrum $p=\uparrow$, the mediated interaction Hamiltonian would be the same as in Eq.~\eqref{eq:effective hamiltonian approximation}, but for a minus sign in the second term, coming from the denominator. This crucial property, which results in the particle spin decoherence being to a large extent removable by the spin echo protocols,\cite{cywinski2009:PRL, cywinski2009:PRB} is thus preserved in the presence of the spin-orbit coupling.

\subsection{Effective Hamiltonian symmetry and magnitude of the spin-non-conserving interactions}

For electrons, we get from Eq.~\eqref{eq:Knight field electrons}
\be
\begin{split}
\tilde{\bf A}^n = R^{-1}_{\boldsymbol{\mathcal{B}}^n} [{\bf A}^n] = \epsilon_p^n \, R^{-1}_{\boldsymbol{\mathcal{B}}^n} \circ R_{U_n}[\langle  J_G | \, {\bf J} \, |J_p \rangle],
\end{split}
\label{eq:electron A vector}
\ee
where we denoted the position dependent energy 
\be 
\epsilon_p^n=-\beta \Phi_G^*({\bf R}_n)\Phi_p({\bf R}_n) \sim -\beta/V.
\ee
Consider first that the magnetic field is small such that the total effective field in Eq.~\eqref{eq:total effective field} is dominated by the Knight field, rather than the external field. In other words, the local impurity quantization axis is collinear with the local particle spin direction. Then $R_{\boldsymbol{\mathcal{B}}^n}=R_{U_n}$ and we get from Eqs.~\eqref{eq:effective hamiltonian final} and \eqref{eq:electron A vector} the effective Hamiltonian in the following form
\begin{equation}
\begin{split}
H_{\rm eff} = & \sum_n \mathcal{B}^n \tilde{I}^n_z
+\sum_{p\in \uparrow} \sum_{n,m}  
\frac{\epsilon_G^n\epsilon_p^m}{E_G - E_p } \tilde{\bf I}^n_z \tilde{\bf I}^m_z \\
&+
\sum_{p\in\downarrow} \sum_{n,m}  
\frac{\epsilon_G^n\epsilon_p^m} {E_G - E_p } \left( \tilde{\bf I}^n_- \tilde{\bf I}^m_+ +\tilde{\bf I}^m_- \tilde{\bf I}^n_+\right)/2.
\end{split}
\label{eq:effective hamiltonian electron 1}
\end{equation}
We have split the summation over the particle excited states into those with the same, and the opposite spin as is the spin of the ground state, corresponding to the second, and the third term in Eq.~\eqref{eq:effective hamiltonian electron 1}, respectively. Equation \eqref{eq:effective hamiltonian electron 1} makes it clear that there is a conserved quantity even in the presence of spin-orbit coupling, though it is neither the energy nor the total spin along any axis; it is the number of impurity spins locally aligned with the particle spin, equal to $\sum_n \tilde{I}_z^n$. This result is very general, as it is based only on the form of the spin-orbit coupling, which gives a single unitary operator $U$ for the whole particle spectrum. Restricting to the lowest excited state, as in Eq.~\eqref{eq:effective hamiltonian approximation}, we get the standard result\cite{shenvi2005:PRB,yao2006:PRB}
\begin{equation}
\begin{split}
H_{\rm eff} = & \sum_n \mathcal{B}^n \tilde{I}^n_z -
\sum_{n,m}  
\frac{\epsilon_G^n\epsilon_G^m} {E_z} \left( \tilde{\bf I}^n_- \tilde{\bf I}^m_+ +\tilde{\bf I}^m_- \tilde{\bf I}^n_+\right)/2,
\end{split}
\label{eq:effective hamiltonian approximation electron}
\end{equation}
generalized to include the effects of the spin-orbit coupling. 

For the electronic case, we are, however, more interested in a different regime, where a finite magnetic field breaks the above discussed symmetry and sets a global quantization axis for impurities, so that the Zeeman energy dominates in the total effective field in Eq.~\eqref{eq:total effective field}. We then have $R_{\boldsymbol{\mathcal{B}}^n}\approx \mathds{1}$ and $\tilde{\bf I}\approx {\bf I}$. Equation \eqref{eq:electron A vector} can be then evaluated explicitly, using Eq.~\eqref{eq:transformation U}. Instead, we estimate the effects of weak spin-orbit coupling, which guarantees that $R_m \ll l_{so}$, by expanding the rotation operator up to the first order as
\be
U({\bf R}_m) \approx \mathbb{I} + O(r_{m}/l_{so}).
\ee
The pairwise interaction in the effective Hamiltonian then appear in combinations such as [see Eq.~\eqref{eq:second order detail} in Appendix \ref{app:A vectors}]
\be
\tilde{I}^n_+ \tilde{I}^m_- + \gamma \tilde{I}^n_+ \tilde{I}^m_+ + \gamma^\prime \tilde{I}^n_+ \tilde{I}^m_z+\ldots,
\label{eq:spin breaking}
\ee
where $\gamma,\gamma^\prime = O[l/l_{so}]$. This is the most important message for the electron case, that the spin-orbit coupling results in the spin-non-conserving interactions in the impurity ensemble, which are, compared to the spin-conserving ones, suppressed by a position dependent factor of the order of the ratio of the confinement and spin-orbit lengths.

We now turn attention to a hole dot, taking the lowest state in the heavy hole subband as the ground state $G=3/2,00,0$. The closest excited state, which gave by far the dominant contribution in the electronic case, is the spin opposite heavy hole state $p=-3/2,00,0$. The corresponding vectors ${\bf A}$ scale as (see Appendix \ref{app:A vectors} for full expressions)
\be
\tilde{A}_+^n \sim \epsilon_p^n O(\lambda_0^2),\qquad \tilde{A}_-^n \sim \epsilon_p^n\lambda_0,\qquad \tilde{A}_z^n \sim \epsilon_p^n\lambda_0.
\label{eq:A vector hole 1}
\ee
To quantify the prefactor in the second order term, $\tilde{A}_+^n$, we would have to go to the next order in the perturbation expansion of the wavefunctions. However, this is not necessary as this term does not enter anywhere in the subsequent discussion. We conclude from Eq.~\eqref{eq:A vector hole 1} that the spin-conserving interactions mediated by the lowest heavy hole excited state are proportional to the second power of parameters $\lambda$ [through terms such as $\tilde{A}_-\tilde{A}^*_-\tilde{I}_+\tilde{I}_-$], the same as the spin-non-conserving ones [e.g. $\tilde{A}_-\tilde{A}^*_z\tilde{I}_+\tilde{I}_z$]. This is a drastic difference to the electron case, where the spin-conserving interactions dominate.

Let us now consider the light hole subband. Taking $p=1/2,00,0$, we get (see Appendix \ref{app:A vectors})
\be
\tilde{A}_+^n \sim \epsilon_p^n,\qquad \tilde{A}_-^n \sim \epsilon_p^n\lambda_0,\qquad \tilde{A}_z^n \sim \epsilon_p^n\lambda_1^\prime.
\label{eq:A vector hole 2}
\ee
The light hole excited state does mediate spin-conserving impurity interactions [through $\tilde{A}_+\tilde{A}^*_+\tilde{I}_-\tilde{I}_+$]. Compared to these, the leading spin-non-conserving term [$\tilde{A}_+\tilde{A}^*_z\tilde{I}_-\tilde{I}_z$] is suppressed linearly in $\lambda$. The energy denominator in the effective Hamiltonian is of the order of 100 meV for the light hole states (typical light-heavy hole band offset) versus a few meV offset of the lowest heavy hole excited state. For our parameters, this energy penalty is almost exactly compensated by much larger matrix elements for the spin-non-conserving interactions and more than compensated for the spin-conserving ones. We conclude that the spin-alike light hole state is the most efficient mediator of the spin-conserving interactions in the impurity ensemble, and rather efficient mediator of the spin-non-conserving ones. As a direct consequence, and unlike for electrons, the decoherence induced by the hole mediated evolution of the impurity bath will not be removed by the hole spin echo.

\section{Phonon induced spin relaxation of the impurity bath}

We now use the results of the previous section to calculate how fast the impurity ensemble spin relaxes. The first and the second term of the effective Hamiltonian, Eq.~\eqref{eq:effective hamiltonian final}, induces flip of a single impurity and a pair of impurities, respectively. For the former, terms with in-plane components of $\tilde{\bf I}$, while for the latter terms such as the last two terms in Eq.~\eqref{eq:spin breaking} are required for spin-non-conserving transitions. As the initial and final state energies differ, in general, we consider that the transition is assisted by phonons, which provide for the energy conservation. 

We consider several possible mechanisms how phonons can couple to the impurity bath and make order of magnitude estimates for the resulting relaxation rates. We find that the most efficient relaxation is due to the piezoelectric field spatially shifting the particle, leading to a $\mu$s relaxation time for Mn spins. It is known that phonons are ineffective in relaxing nuclear spins,\cite{abragam} still we evaluate the resulting rates also for electrons, as we treated electrons and holes on the same footing, the derived formulas apply for both. We find a $10^{11}$ s relaxation time for nuclear spins.

\subsection{Particle-phonon interactions}

The phonon-impurity interaction Hamiltonian $H_i$ is in general a function of the local lattice deformation due to the presence of acoustic phonons,
\be
\delta {\bf R} = {\rm i}  \sum_{{\bf Q}\lambda} \sqrt{ \frac{\hbar}{2 V_0 \rho \omega_{{\bf Q}\lambda}} } {\bf e}_{{\bf Q}\lambda} {\rm e}^ { {\rm  i} {\bf Q}\cdot {\bf R} } \left( a_{{\bf Q}\lambda} + a^\dagger_{-{\bf Q}\lambda} \right).
\label{eq:lattice shift}
\ee
Here the phonon wavevector is ${\bf Q}$, polarization is $\lambda$ (one longitudinal $\lambda=l$ and two transversal ones $\lambda=t_{1,2}$) with $ {\bf e}_{{\bf Q}\lambda}$ a real unit vector ($ {\bf e}_{{\bf Q}\lambda}=- {\bf e}_{-{\bf Q}\lambda}$), $V_0$ is the crystal volume, $\rho$ is the material density, $\hbar \omega_{{\bf Q}\lambda}=\hbar c_\lambda Q$ is the phonon energy, $c_\lambda$ is the phonon velocity, and $a^\dagger_{{\bf Q}\lambda}$ is the phonon creation operator.

In a polar material, such as GaAs, the lattice deformation is accompanied by a piezoelectric field, which is the gradient of the following potential 
\be \begin{split}
V_{PZ} &= -{\rm i}\, \Xi \sum_{{\bf Q}\lambda} \sqrt{ \frac{2\hbar}{ V_0 \rho \omega_{{\bf Q}\lambda} } }  \frac{1}{Q^2} {\rm e}^ { {\rm  i} {\bf Q}\cdot {\bf R} } \left( a_{{\bf Q}\lambda} + a^\dagger_{-{\bf Q}\lambda} \right) \times \\&
\times
\Big (Q_x Q_y ({\bf e}_{{\bf Q}\lambda} )_z +
Q_z Q_x ({\bf e}_{{\bf Q}\lambda} )_y+Q_y Q_z ({\bf e}_{{\bf Q}\lambda} )_x \Big)  
,
\end{split}
\label{eq:piezo}
\ee
with $\Xi$ the piezoelectric constant.

In addition to the previous, the deformation of the lattice shifts the electronic bands, quantified as the deformation potential $V_{DP} =-\sigma\, {\rm div}\, \delta {\bf R}$, which thus reads
\be
\begin{split}
V_{DP} = \sigma \sum_{{\bf Q}} \sqrt{ \frac{\hbar}{2 V_0 \rho \omega_{{\bf Q}l}} } Q \,  {\rm e}^ { {\rm  i} {\bf Q}\cdot {\bf R} }
\left( a_{{\bf Q}l} + a^\dagger_{-{\bf Q}l} \right),
\end{split}
\label{eq:deformation potential}
\ee
with $\sigma$ the deformation potential constant.

In what follows we will see that a relative shift of the impurity and the particle, which we denote by ${\bf d}$, will induce impurity-phonon coupling, leading to the impurity spin relaxation. Since impurities are tied to atoms, the phonon displacement is obviously such a relative shift ${\bf d}=\delta{\bf R}$, which we call ``geometric''. However, the phonon induced electric fields ${\bf E}$ also lead to shifts. Namely, adding the potential of an in-plane field to that in Eq.~\eqref{eq:potential} amounts to a shift of the quantum dot position by ${\bf d}=e{\bf E} l^2 / \hbar \omega$ (electrically induced shifts along the perpendicular direction are much smaller, as the wavefunction is much stiffer along z due to stronger confinement). If the particle follows these potential changes adiabatically, which we assume, such a shift is equivalent to the shift of the impurities, which are fixed to the lattice, by $-{\bf d}$. Since the phonon electric fields are proportional to the displacement $\delta{\bf R}$, we can write a general expression
\be
|{\bf d}| \sim \alpha |\delta {\bf R}|,
\label{eq:alpha}
\ee
with a dimensionless factor $\alpha$. For the geometric shift mechanism $\alpha=1$ by definition. For the piezoelectric field, comparing Eqs.~\eqref{eq:lattice shift} and \eqref{eq:piezo},  we get
\be
\alpha  = 2 Q l  \frac{\Xi l}{\hbar \omega}.
\label{eq:alpha 2}
\ee
Finally, the deformation potential gives
\be
\alpha =  \sigma (Q l)^2 / \hbar \omega.
\ee
We specify the dimensionless factor $\alpha$ in Table \ref{tab:alpha}. As it enters the relaxation rates in the second power (as we will see), we can immediately quantify the relative importance of the three considered channels. Piezoelectric field is the most effective, for both electron and hole cases, inducing shifts almost two orders of magnitude larger than the geometric shift. The electric field from the deformation potential is comparable to the piezoelectric for holes, and much smaller for electrons, which are deeply in the long phonon wavelength limit, $Q l \ll 1$. We note that the geometric shifts will be in fact somewhat more effective than it seems from the table, as they may (unlike the electric fields) shift the wavefunction along the perpendicular direction. This results in an effective enhancement of $\alpha$ by a factor of $\pi l/w$, which, however, is not large enough to change the order of importance following from the Table.

\begin{table}
\begin{tabular}{|c|ccc|}
\hline
$\alpha$ & piezoelectric & deformation & geometric\\
\hline
electron & 46 & 0.0038 &1 \\
hole &38 & 17 &1 \\
\hline
\end{tabular}
\caption{Values for the dimensionless coefficient $\alpha$, the ratio of the induced impurity shift versus the phonon displacement for various shift mechanisms (columns) and particles (rows). Parameters from Appendix \ref{app:parameters} were used (GaAs and ZnTe for the electron and hole case, respectively), the phonon wavevector for electronic case was specified choosing $B=1$ T.}
\label{tab:alpha}
\end{table}

We will calculate the relaxation rate $\Gamma$ due to a general phonon-impurity interaction $H_i$ by the Fermi's golden rule. For a given phonon polarization $\lambda$ it reads
\be
\Gamma
= \frac{2\pi}{\hbar} \sum_{{\bf Q}} |\langle \mathcal{I}^\prime | H_i | \mathcal{I} \rangle |^2  \delta(E_\mathcal{I} - E_{\mathcal{I}^\prime} - \hbar \omega_{{\bf Q}\lambda})N_Q,
\ee
where $\mathcal{I}$ and $\mathcal{I}^\prime$ denote the initial and final state of the impurities, and we are interested in transitions where these two states differ in spin. The phonon occupation factor $N_Q=n_Q+1$, and $N_Q=n_Q$, if the energy of the initial state is larger, and smaller than the final state, respectively, with $n_Q=1/[\exp(\hbar \omega_{{\bf Q}\lambda}/k_BT)-1]$. The energy conservation fixes the phonon wavevector magnitude to $|E_\mathcal{I} - E_{\mathcal{I}^\prime}| = \hbar c_\lambda Q$, by which we get
\be
\Gamma = \frac{V_0Q^2}{\pi \hbar^2 c_\lambda} N_Q\, \overline{| \langle \mathcal{I}^\prime | H_i | \mathcal{I} \rangle |^2}.
\ee
The overline denotes the angular average,  
\be
\overline {f({\bf Q})} = (1/4\pi) \int {\rm d}\Omega\, f({\bf Q}),
\label{eq:angular average}
\ee
over directions of the phonon wavevector ${\bf Q}$.

\subsection{Spin-phonon coupling due to impurity shift}

 Assuming the shifts are small, we get the impurity-phonon coupling as 
\be
H_i = -\sum_n  {\bf d} \cdot \frac{\partial H_{\rm eff}}{\partial {\bf R}}  |_{{\bf R}={\bf R}_n}.
\label{eq:impurity-phonon}
\ee
To calculate the spatial derivative of the effective Hamiltonian, Eq.~\eqref{eq:effective hamiltonian final}, it is easier to first evaluate the derivative of vectors $\boldsymbol{\mathcal{B}}$ and ${\bf A}$ in the original coordinate system which does not depend on the position and then to transform them into the locally rotated coordinates. 

Let us start with the first term of the effective Hamiltonian, Eq.~\eqref{eq:effective hamiltonian final}. The finite derivative of the total field, Eq.~\eqref{eq:total effective field}, is due to the spatial dependence of the Knight field, 
\be
({\bf d} \cdot \partial_{\bf R}) \boldsymbol{\mathcal{B}}^n =
({\bf d} \cdot \partial_{\bf R}) {\bf K}^n,
\label{eq:K derivative}
\ee 
transversal components of which give the spin increasing transition rate for impurity $n$ as
\be
\Gamma^{(1)} =
\frac{V_0Q^2}{\pi \hbar^2 c_\lambda} N_Q \overline{[ {\bf r^\prime}_- \cdot ({\bf d}\cdot \partial_{\bf R}) {\bf K} ]^2} |\langle I^{n+1} | \tilde{I}^n_+ | I^n \rangle |^2,
\label{eq:first rate}
\ee
and an analogous expression follows swapping the subscripts $\pm$ for a spin decreasing transition. Our goal is an order of magnitude estimate for the rates as the one in Eq.~\eqref{eq:first rate}, which allows us to simplify: We replace the spin dependent matrix elements of the raising/lowering operators by
\be
|\langle I^{n\pm 1} | \tilde{I}^n_\pm | I^n \rangle |^2 = I(I+1)-I^n(I^n\pm 1) \sim I^2,
\ee 
choose the direction of the phonon wavevector that gives the highest contribution, instead of performing the angular average in Eq.~\eqref{eq:angular average}, and denote 
\be
\nabla\tilde{\bf K}_- \equiv \delta^{-1}\, \overline{ {\bf r^\prime}_- \cdot ({\bf d} \cdot \partial_{\bf R}) {\bf K} }.
\ee
Finally, we use Eq.~\eqref{eq:alpha} and $\delta R \sim \surd \hbar/2V_0\rho c_\lambda Q$ to write
\be
\Gamma^{(1)} \sim \frac{\alpha^2 I^2 N_Q Q}{2\pi \hbar c_\lambda^2 \rho} (\nabla\tilde{\bf K}_-)^2 ,
\label{eq:aux5}
\ee
a general form for the relaxation rate estimate, which we a few lines below evaluate for specific cases.

Let us start with the electronic case. The transition energy is dominated by the external field $\hbar c_\lambda Q \approx |g\mu_N B|$ and is much smaller than the thermal energy, so that $N_Q\approx  k_BT / \hbar c_\lambda Q$. We evaluate the derivative of the Knight field in Appendix \ref{app:K vectors}, see Eq.~\eqref{eq:Knight field electrons derivative 2}, getting
\be
\nabla \tilde{\bf K}_\pm \sim (\beta/V) l_{\rm so}^{-1}.
\label{eq:transversal K}
\ee
For the dominant piezoelectric mechanism we get
\be
\Gamma^{(1)} \sim \frac{9 I^2}{2\pi^3} \frac{k_B T (g \mu_N B)^2}{\hbar E_{\rm pz}^2} \frac{l^4}{w^2 l_{\rm so}^2} \label{eq:gamma1}.
\ee
where the energy $E_{\rm pz} = \sqrt{\hbar^7 c_\lambda^5 \rho}/\Xi m \beta$ is a material constant. Evaluating parameters of GaAs, we get a minuscule rate $\Gamma^{(1)} \sim 2 \times 10^{-11} $s$^{-1}$, choosing transversal phonons, external field 1 Tesla and temperature 1 Kelvin.

We now turn to holes. The transferred energy is now given by the Knight field, rather than the external field,  $|E_\mathcal{I} - E_{\mathcal{I}^\prime}| \sim J \beta/V$ and we again consider a high temperature limit, $k_B T\geq |E_\mathcal{I} - E_{\mathcal{I}^\prime}|$. As we show in Appendix \ref{app:K vectors}, the formula in Eq.~\eqref{eq:transversal K} is changed into 
\be
\nabla \tilde{\bf K}_\pm \sim (\beta/V) (\sqrt{3}\lambda_1/ l),
\ee
showing that quantities of the form $l/\lambda$ can be seen as an effective spin-orbit length for holes. Equation ~\eqref{eq:gamma1} can be then used putting for the ``spin-orbit length'' the one just described and replacing the external Zeeman energy by the Knight field. For convenience, we give the relaxation rate explicitly
\be
\Gamma^{(1)} \sim \frac{3^5 I^2 J^2}{2^5\pi^4} \frac{k_B T}{\hbar} \frac{ \beta^2}{E_{\rm pz}^2} \frac{\lambda_1^2}{w^4 l^2},
\ee
which for ZnTe parameters and temperature $10$ K yields a modest rate $0.35\,\mu$s$^{-1}$. 
The rate is second order in the ``spin-orbit strength'' and, unlike for electrons, grows very fast upon making the dot smaller, since now the effect of Knight field, inversely proportional to the dot volume, dominates over the effect of the shift being larger for softer dot potential.

Comparing the two terms of the effective Hamiltonian, Eq.~\eqref{eq:effective hamiltonian final}, and using the results of Appendix \ref{app:A vectors}, the pair-wise spin transition mediated by the lowest heavy hole excited state relates to the single spin-flip rate by
\be
N\Gamma^{(2)}_{p=3/2} /  \Gamma^{(1)} \sim  N I^2 \left( 2 \frac{\lambda_0^2}{\lambda_1} \frac{\beta/V}{E_\downarrow-E_\uparrow} \right)^2,
\ee
which can be evaluated as $10^{-2}$. Similarly, we get for the mediation through the light hole state
\be
N\Gamma^{(2)}_{p=1/2} / \Gamma^{(1)} \sim N \left( \frac{3\lambda_1^\prime}{4\lambda_1} \frac{\beta/V}{\Delta_{lh}} \right)^2,
\label{eq:aux6}
\ee
where a much larger energy offset, $\Delta_{lh}$, is partially compensated by a larger matrix element. Note that the number of pairs available for a flip is of the order of $N$ times larger than the spins themselves, so that when comparing the first and the second order rate, the latter should be multiplied by $N$, what we did in the previous two equations. 

Just for completeness we note that for electrons a similar relation between the first and second order rates holds,
\be
\Gamma^{(2)} \sim \Gamma^{(1)} \left(\frac{\beta / V}{E_\downarrow-E_\uparrow}\right)^2 I^2 \ll \Gamma^{(1)},
\label{eq:gamma2}
\ee
but the ratio is much smaller, at the external field of 1 Tesla by 10 orders of magnitude. 

We now consider additional mechanisms of the phonon-impurity couplings, through which impurity spin relaxation may arise.

\subsection{Valence band shifts}

The phonon induced lattice compression changes the bandstructure -- the bands are shifted. Shifts different for the bands of the particle ground and mediating excited state $p$ result in the impurity-phonon coupling through the second term of Eq.~\eqref{eq:effective hamiltonian final}, by changing the denominator by $\Delta V_{DP}$. As the two states have to belong to different bands, such a coupling may arise only for the case of holes and takes the form
\be
H_i \sim \frac{\Delta V_{DP}}{\Delta_{lh}} H_{\rm eff},
\label{eq:band shift}
\ee
expanding the effective Hamiltonian up to the lowest order in the band shift difference $\Delta V_{DP} =-(\sigma_{hh}-\sigma_{lh}) {\rm div} \delta {\bf R}$, which is the difference of the deformation potentials for the heavy and light hole valence bands. We can thus relate the band shift mechanism to the position shift one, comparing Eq.~\eqref{eq:impurity-phonon} with Eq.~\eqref{eq:band shift}. We find that the latter is described by an effective constant $\alpha$
\be
\alpha_{\rm eff} = ( \sigma_{hh} - \sigma_{lh} ) \sigma Q l / \Delta_{lh}.
\label{eq:alpha eff}
\ee
If we estimate the potential difference by the typical value of the potential itself, $(\sigma_{hh}-\sigma_{lh}) \sim 5$ eV, which is likely an overestimated value, we get $\alpha_{\rm eff}=17$, so that the coupling through band shifts leads to a rate at most comparable to (and most probably much smaller than) that described by Eq.~\eqref{eq:aux6}.

\subsection{Renormalization of the spin-orbit length}

Phonon induced renormalization of band offsets influences the spin-orbit couplings. This is evident from the expressions for the coefficients $\lambda$ which are inversely proportional to the light-heavy hole offset $\Delta_{lh}$, see Eqs.~\eqref{eq:lambdas}. Even though the coupling in the form of the constant $\alpha$ is described by the same formula as in Eq.~\eqref{eq:alpha eff}, the substantial difference is that now the phonon-impurity coupling arises also through the first term of the effective Hamiltonian because fluctuations in spin-orbit fields induce fluctuations in the Knight field. The corresponding $\alpha_{\rm eff}$ is the one given in Eq.~\eqref{eq:alpha eff} and the relaxation rate is that in Eq.~\eqref{eq:aux5}, so that it does not exceed the rate due to the piezoelectric shift mechanism. For the case of electrons, the effective constant follows in an analogous form, 
\be
\alpha = ( \sigma_e - \sigma_h ) \sigma Q l_{\rm so} / \Delta,
\ee
where $\Delta$ is of the order of the conduction-valence band offset, which enters the definition of the spin-orbit couplings $1/l_{\rm so}$. The numerical value for $\alpha$ is much less than one even taking $(\sigma_e - \sigma_h) \sim \sigma_e$, so that this channel is negligible with respect to the geometrical shifts.

\subsection{Phonon induced spin-orbit interactions}

Finally, we estimate the influence of spin-phonon coupling through an additional spin-orbit interaction arising in the presence of a phonon-originated electric field. We assume the new spin-orbit interaction strength relates to the one we considered in Sec.~II, referred to as ``old'', through the ratio of the internal (interface) electric field $E_{\rm int}$ and the phonon induced electric field, the latter given as the gradient of the appropriate potential, Eqs.~\eqref{eq:piezo} or \eqref{eq:deformation potential}. Since the phonon-induced spin-orbit interaction (``new'') arises from an electric field, it is of the Rashba functional form, and one can take its effects to be additive to the ``old'' one. For electrons, this means 
$U_{tot}= U_{old} U_{new} \approx (\mathbb{I}+O_{old}) (\mathbb{I} + O_{new}) \approx \mathbb{I} + (O_{old}+O_{new})$, which amounts to additive inverse of the spin-orbit lengths $ 1/l_{so} \to 1/l_{so}  + 1/ l_{ph-so}$. By inspecting Eq.~\eqref{eq:Knight field electrons derivative}, we estimate the effective interaction to be described again by Eq.~\eqref{eq:aux5} with the constant
\be
\alpha_{\rm eff} = \frac{\Xi Q l}{eE_{\rm int}},
\label{eq:alpha eff 2}
\ee
for the piezoelectric phonon field (a much smaller deformation field corresponds to the numerator replaced by $\sigma Q^2 l$). The numerator evaluates to $10^6$ V/m, which is not supposed to be much higher than the internal field, so that again, we find that this mechanism is less important than the one due to the piezoelectric shift. We come to a similar conclusion for holes (though we do not show calculations in detail here). Namely, even though the phonon induced piezoelectric field is much stronger, of the order of $10^8$ V/m, the spin-orbit terms in Kohn-Luttinger Hamiltonian are less effective in inducing light-heavy hole mixing than the terms we considered explicitly in previous sections (see Appendix \ref{app:alternative mixing}).

\section{Conclusions}

We have analyzed the interactions within an ensemble of impurity spins, which are mediated by a confined spin-orbit coupled particle. We have considered two physical systems where such mediated interactions are of great importance: III-V (GaAs) electronic lateral quantum dot with the impurities being nuclear spins, and II-VI (ZnTe) self-assembled Mn-doped quantum dot populated by a hole. Our focus has been on the consequences of the spin-orbit coupling on the character of the mediated interactions. 

We have derived an effective Hamiltonian for the impurity ensemble, treating the particle-impurity interaction perturbatively. The form of this Hamiltonian allowed us to quantify the degree to which the conservation of impurities spin is broken in the presence of the spin-orbit coupling of the particle. We have found that for the electron case, the spin-non-conserving terms are suppressed relative to the spin-conserving ones by a small factor, the ratio of the confinement length and the spin-orbit length. The lowest electron excited state is the most effective mediator, what results in a decoherence being removable by the electron spin echo even in the presence of the spin-orbit coupling. In the case of holes, the spin-conserving interactions are most efficiently mediated by the lowest light hole state, while the spin-non-conserving ones by the lowest heavy hole state. The induced decoherence is then not removable by a hole spin echo anymore. 

As a direct application of the derived effective Hamiltonian, we have calculated the rates of a phonon-assisted impurity spin relaxation, which arises only in the presence of the spin-orbit coupling in the particle Hamiltonian. We have considered several coupling mechanisms, by which impurity spins couple to phonons. We have found that the most effective is the piezoelectric field induced shift of the particle wavefunction, with a typical relaxation time of 1 $\mu$s in a 10 nm self-assembled strain-free quantum dot. The rate grows upon making the dot smaller, or diminishing the heavy-light hole splitting by strain, possibly to nano seconds for reasonable dot parameters.

While we have focused on a single particle, our analysis of the spin-non-conserving mechanisms already provides insights in the magnetic ordering in quantum dots with multiple occupancy.  For example, the spin-non-conserving mechanisms are the key in understanding the prediction of piezomagnetic quantum dots or, equivalently, nonlinear magneto-electric effects.\cite{abolfath2008:PRL} Changes in the shape of lateral confinement  (from circular to elliptical) controlled by the pair of the gate electrodes have been demonstrated experimentally to alter the particle configuration from vanishing to finite spin in nonmagnetic quantum dots.\cite{austing1999:PRB} This principle provides intriguing possibilities for the control of magnetic ordering in dots with added Mn impurities where such changes in the particle spin, through exchange interaction, would reversibly control the magnetic ordering of the nearby Mn spins.\cite{abolfath2008:PRL} The characteristic time scale for the related magnetic polaron formation, similar to the better studied quantum well structures, should be largely determined by the anisotropic spin-spin interactions of impurities which explicitly do not conserve the total spin of Mn atoms.\cite{yakovlev, dietl1995:JMMM, henneberger, yakovlev2}

Beyond epitaxial dots that we have presently examined, recent experimental advances in colloidal quantum dots warrant also future considerations. Typically they are easily synthesized II-VI materials, such as ZnTe, ZnSe, CdS, and CdSe,\cite{klimov2007:ARPCH,scholes2008:AFM} which offer a large size-induced tunability of the transition energies  and long spin decoherence times.\cite{stern2005:PRB} Magnetic doping\cite{beaulac2008:AFM} of these colloidal dots provide an opportunity for a versatile control of magnetic order as well as lead to robust magnetic polaron formation with effective internal magnetic field approaching 
100 T.\cite{yakovlev,beaulac2008:AFM,bussian2009:NM,ochsenbein2009:NN,zutic2009:NN,viswanatha2011:PRL}

\section{Acknowledgements}

We would like to thank Rafal Oswaldowski for many discussions, which substantially contributed to this work. P.S. would also like to acknowledge useful discussions with Uli Zuelicke. This work was supported by EU project Q-essence, meta-QUTE ITMS NFP 26240120022, CE SAS QUTE, SCIEX, DOE-BES, ONR, and DFG SFB 689.

\appendix

\section{Kohn-Luttinger Hamiltonian perturbative eigenstates}

\label{app:KLH}

Here we derive hole eigenstates in the lowest order perturbation theory. We neglect the influence of the conduction and spin-orbit split-off subbands and consider only the light ($J=\pm1/2$) and heavy ($J=\pm3/2$) holes in the Kohn-Luttinger Hamiltonian,\cite{winkler} $H_{JJ^\prime}$. The diagonal elements are
\begin{equation}
 \begin{split}
H_{\pm 3/2,\pm 3/2} &= -\frac{\hbar^2}{2m_0} \left[ k_z^2(\gamma_1-2\gamma_2) +(k_x^2+k_y^2)(\gamma_1+\gamma_2)  \right],\\
H_{\pm 1/2,\pm 1/2} &= -\frac{\hbar^2}{2m_0} \left[ k_z^2(\gamma_1+2\gamma_2) +(k_x^2+k_y^2)(\gamma_1-\gamma_2)  \right],
\end{split}
\label{eq:LH diagonal}
\end{equation}
with $\hbar{\bf k}={\bf P}$ the hole momentum operator, $m_0$ the free electron mass, and $\gamma_1$, $\gamma_2$, and $\gamma_3$ (below) the Luttinger parameters. Together with the in-plane $V({\bf r})$ and heterostructure $V_z(z)$ confinement potentials, assumed to be those in Eq.~\eqref{eq:potential}, the kinetic terms in Eq.~\eqref{eq:LH diagonal} define the dot unperturbed eigenstates (normalization omitted)
\be
\Phi^J_{nm,k} = r^{|m|} e^{-r^2/2l_J^2} L_n^{|m|} (r^2/l_J^2) e^{{\rm i} m\phi} \sin ( k\pi z/w ).
\ee
We used cylindrical coordinates $(r, \phi, z)$, $L_n^m$ are the associated Laguerre polynomials, $l_J$ the in-plane confinement length (which differs for heavy and light holes due to their different masses) and $w$ the heterostructure width. The Fock-Darwin states are labelled by the principal and orbital quantum numbers $n$ and $m$, and $k$ labels excitations in the perpendicular potential. The corresponding energies are
\begin{equation} 
E_{J,nm,k}=\hbar \Omega_J (2n+|m|+1)+\frac{\hbar^2}{2 m_J w^2} \pi^2 k^2,
\end{equation} 
where the in-plane excitation energy is parameterized by the mass and confinement length,
\be
\hbar \Omega_J = \frac{\hbar^2} {m_J l_J^2}.
\ee
The in-plane masses are given by Eq.~\eqref{eq:LH diagonal}: $m_{\pm 3/2} \equiv m_{hh}=m_0/(\gamma_1+\gamma_2)$, $m_{\pm 1/2} \equiv m_{lh}=m_0/(\gamma_1-\gamma_2)$. We parameterize the in-plane electrostatic potential choosing a certain value for the heavy hole in-plane excitation energy $E_{3/2, 01,0}-E_{3/2,00,0}=\hbar \Omega_{hh}$, which then specifies the confinement lengths. The light hole excitation energy and confinement length
\be
\Omega_{lh} = \Omega_{hh} (m_{hh}/m_{lh})^{1/2}, \qquad l_{lh} = l_{hh} (m_{hh}/m_{lh})^{1/4}, 
\ee
differ from the corresponding heavy hole quantities due to a different in-plane mass. The energies of the hard-wall eigenstates also differ for heavy and light holes due to different out-of-plane masses, which are $m_0/(\gamma_1 + 2\gamma_2)$, and $m_0/(\gamma_1 - 2\gamma_2)$, respectively. We set $w$ by choosing a certain value for the light-heavy hole splitting $E_{3/2,00,0} - E_{1/2,00,0}=\Delta_{lh}$.

Taking the above eigenstates as the basis, we now perturbatively take into account the off-diagonal elements of the Hamiltonian,
\be
\begin{split}
H_{\pm 3/2, \pm 1/2} & = \pm \frac{\hbar^2}{2m_0} 2\sqrt{3}\gamma_3 k_{\mp} k_z ,\\
H_{\pm 3/2, \mp 1/2} & = \frac{\hbar^2}{2m_0} \sqrt{3} \left[ \gamma_2 (k_x^2-k_y^2) \mp 2{\rm i} \gamma_3 k_x k_y \right] ,\\
H_{\pm 3/2, \mp 3/2} & = 0 = H_{\pm 1/2, \mp 1/2}.
\end{split}
\ee
We employ the non-degenerate perturbation theory
\begin{equation}
|\Psi_{Ja}\rangle = |J\rangle \otimes |\Phi_a^J \rangle + \sum_{J^\prime a^\prime \neq J a} \frac{\langle \Phi_{a^\prime}^{J^\prime} | H_{J^\prime J} | \Phi_a^J\rangle}{E_{Ja}-E_{J^\prime a^\prime}} |J^\prime \rangle \otimes |\Phi_{a^\prime}^{J^\prime} \rangle,
\label{eq:hole 1-st order states}
\end{equation}
where we use the notation introduced below Eq.~\eqref{eq:electronic basis 0}, so that $a$ includes two quantum numbers of the in-plane Fock-Darwin state and one of the perpendicular hard-wall state. To proceed, we neglect high energy excitations $n>0$ and $k>1$ and adopt the axial approximation 
\be
H_{\pm 3/2, \mp 1/2} \approx \frac{\hbar^2}{2m_0} \sqrt{3} \gamma k_{\mp}^2,
\ee
with $\gamma=(\gamma_2+\gamma_3)/2$. With these simplifications, the otherwise infinite sum for $|\Psi_{Ja}\rangle$ simplifies to only a single term for each $J^\prime\neq J$ and can be given explicitly.\cite{bhattacharjee2007:PRB, fischer2010:PRL} We finally get Eq.~\eqref{eq:1-st order ground state} with the admixtures
\begin{subequations}
\begin{eqnarray}
\lambda_1 &=& \frac{\hbar^2}{m_0 l_{lh} w} \frac{\gamma_3 \sqrt{3}\kappa\xi}{\Delta^\ast+\hbar \Omega_{lh}},\\
\lambda_0 &=&  \frac{\hbar^2}{m_0 l_{lh}^2} \frac{\gamma \sqrt{3/2}\kappa}{\Delta_{lh}+2\hbar \Omega_{lh}} ,
\end{eqnarray} 
\label{eq:lambdas}
\end{subequations}
where $\Delta^\ast=E_{1/2,00,1}-E_{3/2,00,0}$ is the z-excited light hole offset and 
\be 
\kappa = \langle \Phi^{hh}_{00,0} | \Phi^{lh}_{00,0} \rangle = \frac{2}{(m_{hh}/m_{lh})^{1/4} + (m_{lh}/m_{hh})^{1/4}},
\ee
is the ground state overlap, which differs from one due to the in-plane mass difference. Finally, the dimensionless matrix element $\xi$ is defined by $\xi= -{\rm i} w \langle 1 | k_z | 0\rangle = 8/3$.

Taking a heavy hole 3/2, the admixture of 1/2 light hole scales as $1 / w l$, costs the in-plane plus perpendicular orbital energy (the latter is even larger than the light-heavy hole offset) and leads to an admixture with a very different z-profile (compared to the main wavefunction component). The admixture of the -1/2 light hole has a smaller numerator, proportional to $1/l^2$, but costs only the in-plane orbital energy (several times smaller than the light-heavy hole offset) and has the same z-profile as the main component.

Along the same lines we get Eq.~\eqref{eq:1-st order ground state 3} with 
\begin{subequations}
\begin{eqnarray}
\lambda_1^\prime &=& \frac{\hbar^2}{m_0 l_{hh} w} \frac{\gamma_3 \sqrt{3}\kappa\xi}{\Delta^{*\prime}+\hbar \Omega_{hh}},\\
\lambda_0^\prime &=&  \frac{\hbar^2}{m_0 l_{hh}^2} \frac{\gamma \sqrt{3/2}\kappa}{\Delta_{lh}-2\hbar \Omega_{hh}},
\end{eqnarray} 
\label{eq:lambdas prime}
\end{subequations}
where $\Delta^{*\prime}=E_{3/2,00,1}-E_{1/2,00,0}$ is the z-excited heavy hole offset.

For completeness, we list the hole Zeeman term\cite{katsaros2011:CM} 
\begin{equation}
H_{hZ}= 2\kappa \mu_B {\bf J}\cdot {\bf B} + 2 q \mu_B \sum_{i=x,y,z} J_i^3 B_i,
\end{equation}
which can be written for the heavy hole subspace as
\begin{equation}
H_{hh,Z}= g_{hh} \mu_B [{\bf J}/3]_z \cdot {\bf B}_z,
\end{equation}
with ${\bf J}/3 \equiv \boldsymbol{\sigma}/2$ the pseudo spin operator and $g_{hh} \equiv 6\kappa +27 q/2 \approx 2$ for GaAs.\cite{fischer2010:PRL}

\section{Spin matrix elements for holes}

\label{app:hole matrix elements}

Here we calculate the matrix elements of the spin operator between perturbative eigenstates of the hole. For the purposes of this appendix, we shorten the expression in Eq.~\eqref{eq:1-st order ground state} introducing $\Phi =\Phi_{00,0}^{hh}$, $a = \lambda_1\Phi^{lh}_{01,1}$ and $b=\lambda_0\Phi^{lh}_{02,0}$ to 
\begin{eqnarray}
|\Psi_{3/2} \rangle &=& \Phi |3/2\rangle + a |1/2\rangle + b|-1/2\rangle,\\
|\Psi_{-3/2} \rangle &=& \Phi^* |-3/2\rangle - a^* |-1/2\rangle + b^*|1/2\rangle.
\end{eqnarray}
We denote $J_\pm = J_x \pm {\rm i} J_y$, so that $J_x = (J_++J_-)/2$ and $J_y = (J_+-J_-)/2{\rm i}$ and since the orbital operator in all the matrix elements is the delta function $\delta({\bf R}-{\bf R}_n)$, all the complex amplitudes below should be evaluated at the position of the particular impurity, e.g., $\Phi\to \Phi({\bf R}_n)$. Listing only the leading order in small quantities $\lambda_{0,1}$, to which $a$ and $b$ are proportional, we have
\begin{subequations}
\begin{eqnarray}
\langle \Psi_{3/2} | J_z | \Psi_{3/2} \rangle &=& (3/2) |\Phi|^2 + O(\lambda^2),\\
\langle \Psi_{3/2} | J_+ | \Psi_{3/2} \rangle &=& \sqrt{3}\Phi^*a + O(\lambda^2),\\
\langle \Psi_{3/2} | J_- | \Psi_{3/2} \rangle &=& \sqrt{3}\Phi a^* + O(\lambda^2),
\end{eqnarray}
\label{eq:ME for J 1}
\end{subequations}
from where Eq.~\eqref{eq:Knight field holes} follows directly.
The time reversal symmetry gives (using $\langle T a| b\rangle = -\langle  a| T b\rangle^*$, and $T^2=-1$)
\begin{equation}
\langle \Psi_{-3/2} | {\bf J} | \Psi_{-3/2} \rangle = -\langle \Psi_{3/2} | {\bf J} | \Psi_{3/2} \rangle,
\end{equation}
so that the spin expectation value changes sign upon inverting the hole spin. To evaluate the off-diagonal element of the Overhauser field and the vectors ${\bf A}$ we need 
\begin{subequations}
\begin{eqnarray}
\langle \Psi_{-3/2} | J_z | \Psi_{3/2} \rangle &=& ab,\\
\langle \Psi_{-3/2} | J_+ | \Psi_{3/2} \rangle &=& 2 b^2,\\
\langle \Psi_{-3/2} | J_- | \Psi_{3/2} \rangle &=& 2\sqrt{3}\Phi b + O(\lambda^2),
\end{eqnarray}
\label{eq:ME for J 2}
\end{subequations}
from where Eq.~\eqref{eq:transversal Overhauser} follows.

Analogously we define short hand notations for the light hole like wavefunctions as $\Phi^\prime = \Phi^{lh}_{00,0}$, $a^\prime = \lambda_1^\prime \Phi^{hh}_{0\overline{1},1}$, and $b^\prime=\lambda_0^\prime \Phi^{hh}_{0\overline{2},0}$, to write
\begin{eqnarray}
|\Psi_{1/2} \rangle &=& \Phi^\prime |1/2\rangle + a^\prime |3/2\rangle - b^\prime|-3/2\rangle,\\
|\Psi_{-1/2} \rangle &=& \Phi^{\prime*} |-1/2\rangle - a^{\prime*} |-3/2\rangle - b^{\prime*}|3/2\rangle.
\end{eqnarray}
The light hole mediated effective interaction is given by
\begin{subequations}
\begin{eqnarray}
\langle \Psi_{3/2} | J_z | \Psi_{1/2} \rangle &=& (3/2) \Phi^* a^\prime + (1/2) a^\ast \Phi^\prime,\\
\langle \Psi_{3/2} | J_+ | \Psi_{1/2} \rangle &=& \sqrt{3}\Phi^*\Phi^\prime+ O(\lambda^2),\\
\langle \Psi_{3/2} | J_- | \Psi_{1/2} \rangle &=& 2 \Phi^\prime b^* + O(\lambda^2).
\end{eqnarray}
\label{eq:ME for J 3}
\end{subequations}
Finally, the coupling through the spin opposite light hole state are given by
\begin{subequations}
\begin{eqnarray}
\langle \Psi_{3/2} | J_z | \Psi_{-1/2} \rangle &=& -(3/2) \Phi^* b^{\prime*}  -(1/2) b^\ast \Phi^{*\prime},\\
\langle \Psi_{3/2} | J_+ | \Psi_{-1/2} \rangle &=& 2a^*\Phi^{\prime*} + O(\lambda^2),\\
\langle \Psi_{3/2} | J_- | \Psi_{-1/2} \rangle &=& -\sqrt{3} b^{\prime*} a^*.
\end{eqnarray}
\end{subequations}

\section{Knight field and rotated coordinates}

\label{app:K vectors}

In this appendix, we evaluate the Knight field, its spatial derivative and the corresponding locally rotated coordinate frame, for the case of electrons and holes.

\subsection{Electron case}

The Knight field of an electron in the ground state is
\be
{\bf K} = -(\beta/2) | \Phi_G({\bf R}) |^2 \vecs{},
\label{eq:Knight field electrons app}
\ee
see Eq.~\eqref{eq:Knight field electrons}. In deriving that, we used the identity
\be
U {\bf J} U^\dagger =  R_U^{-1} [ {\bf J} ],
\label{eq:rotation identity}
\ee
where the unitary operator of spinor rotation
\be
U \equiv \exp(-{\rm i}\, {\bf n} \cdot {\bf J}\, \phi),
\ee
corresponds to a three dimensional rotation around the unit vector ${\bf n}$ by angle $\phi$,
\be
R_{U}= \exp ( -{\rm i} \, {\bf n} \cdot {\bf l}\, \phi),
\ee
with $(l_k)_{mn} = -{\rm i} \epsilon_{kmn}$.

The spatial derivative of the Knight field, which gives the matrix elements for the impurity spin flips in the ''first order`` rates, follows from Eq.~\eqref{eq:Knight field electrons app} as
\be
\nabla {\bf K}  = \left[ \nabla \ln(| \Phi_G({\bf R}) |^2) \right] {\bf K} + \left( \nabla {\bf n}_{\rm so} \right) \times {\bf K}.
\label{eq:Knight field electrons derivative}
\ee
The two terms correspond, respectively, to the position change in the magnitude and direction of vector ${\bf K}$. 

For electrons we consider a regime in which the total field is dominated by the external field. The impurity local coordinate system then coincides with the coordinate frame defined by the external magnetic field, $R_{{\bf \mathcal{B}}^n} \approx \mathds{1}$ and ${\bf \hat{z}^\prime} = {\bf s}_0$. This allows us to estimate the components of $\nabla {\bf K}$ transversal to the local coordinate system as 
\be
{\bf r^\prime}_\pm \cdot \nabla {\bf K} \sim \delta^{-1} (l/l_{\rm so}) K,
\label{eq:Knight field electrons derivative 2}
\ee
which originate in the first term of Eq.~\eqref{eq:Knight field electrons derivative} and where the length $\delta=l$ for a phonon with in-plane polarization vector (${\bf e} \perp {\bf \hat{z}}$) and $\delta=w/\pi$ for a phonon with a polarization vector along the growth direction (${\bf e} \,||\, {\bf \hat{z}}$). The second term of Eq.~\eqref{eq:Knight field electrons derivative} gives a contribution at most as large as the first term, or lower, depending on the phonon polarization.

\subsection{Holes}

The Knight field of a spin 3/2 hole, defined by Eq.~\eqref{eq:Knight field},
\begin{equation}
{\bf K}^n= -\beta \langle \Psi_{3/2} | \delta({\bf R}-{\bf R}_n)\, {\bf J}\, | \Psi_{3/2} \rangle,
\label{eq:aux3}
\end{equation}
follows from Eqs.~\eqref{eq:ME for J 1} as
\be
[ K_x^n, K_y^n, K_z^n] = -\beta [\sqrt{3}\,{\rm Re} (a\Phi^*), \sqrt{3}\,{\rm Im} (a\Phi^*), 3/2\, |\Phi|^2 ],
\ee
given in the main text in Eq.~\eqref{eq:Knight field holes}, here using the notation from Appendix \ref{app:hole matrix elements}, $a=\lambda_1 \Phi_{01,1}^{lh}({\bf R}_n)$ and $\Phi=\Phi_{00,0}^{hh}({\bf R}_n)$. Since for holes we are interested in the zero magnetic field case (by which $\Phi$ is real), the local coordinate frame is defined by a unit vector along the Knight field, ${\bf \hat{z}^\prime} = {\bf K}/K$,
\be
{\bf \hat{z}^\prime} = [\sin\phi\, {\rm Re} (a)/|a|,\sin\phi\, {\rm Im} (a)/|a|, \cos\phi].
\ee
Here $\phi$ is the angle between the original and the rotated z axes, $\cos \phi = {\bf \hat{z}} \cdot {\bf \hat{z}^\prime} = \Phi/\sqrt{3\Phi^2/4+|a|^2}$. We choose the remaining two axes of the rotated coordinate system arbitrarily as  
\be 
{\bf \hat{y}^\prime} = [ -{\rm Im} (a)/|a|, {\rm Re} (a)/|a|, 0],
\ee
and
\be
{\bf \hat{x}^\prime} = {\bf \hat{y}^\prime} \times {\bf \hat{z}^\prime} = 
 [\cos \phi \,{\rm Re} (a)/|a|, \cos \phi \,{\rm Im} (a)/|a|, -\sin \phi].
\ee
Using the projectors into the transversal plane of the local coordinate system, ${\bf r^\prime}_\pm$, one can evaluate the amplitudes of the spin-non-conserving terms. As an auxiliary result, we note that for any real vector ${\bf v}$ we have
\be
{\bf r^\prime}_\pm \cdot {\bf v} = v_+ \frac{a^*}{|a|} \frac{\cos\phi\pm1}{2}+v_- \frac{a}{|a|} \frac{\cos\phi\mp1}{2}-v_z\sin\phi.
\label{eq:aux4}
\ee
For example, close to the dot center, the inequality $\Phi \gg |a|$ gives $\sin\phi\approx |a|/\sqrt{3}/2\Phi$, $\cos\phi\approx 1$, $|a|\sim\lambda_1\Phi$, and the Knight field derivative follows as 
\be
{\bf r^\prime}_\pm \cdot \nabla {\bf K} \sim \delta^{-1} \sqrt{3} \lambda_1 \beta |\Phi_{00,0}^{hh}|^2,
\ee
where, again, the length $\delta$ depends on the direction of the phonon polarization vector, $\delta=l$ for ${\bf e_Q} \perp {\bf \hat{z}}$, and $\delta=w/\pi$ for ${\bf e_Q} \,||\, {\bf \hat{z}}$.

\section{Interactions in the impurity ensemble: vectors ${\bf A}$ and the effective Hamiltonian terms}

\label{app:A vectors}

The second order Hamiltonian for a given impurity pair $n,m$ equals $1/(E_G-E_p)$ times the following expression 
\be
\begin{split}
 &({\bf A}^n\cdot {\bf I}^n) ({\bf A}^{m} \cdot {\bf I}^m)^\dagger + ({\bf A}^m\cdot {\bf I}^m)( {\bf A}^{n} \cdot {\bf I}^n)^\dagger=\\
& \qquad I^n_z I^m_z (A^n_z A^{m*}_z + A^m_z A^{n*}_z) +\\
& \qquad I^n_+ I^m_- (A^n_- A^{m*}_- + A^m_+ A^{n*}_+)/4 + \\
& \qquad I^n_- I^m_+ (A^n_+ A^{m*}_+ + A^m_- A^{n*}_-)/4 + \\
& \qquad I^n_z I^m_+ (A^n_z A^{m*}_+ + A^m_- A^{n*}_z)/2 + \\
& \qquad I^n_+ I^m_z (A^n_- A^{m*}_z + A^m_z A^{n*}_+)/2 + \\
& \qquad I^n_z I^m_- (A^n_z A^{m*}_- + A^m_+ A^{n*}_z)/2 + \\
& \qquad I^n_- I^m_z (A^n_+ A^{m*}_z + A^m_z A^{n*}_-)/2 + \\
& \qquad I^n_+ I^m_+ (A^n_- A^{m*}_+ + A^m_- A^{n*}_+)/4 + \\
& \qquad I^n_- I^m_- (A^n_+ A^{m*}_- + A^m_+ A^{n*}_-)/4.
\end{split}
\label{eq:second order detail}
\ee
If the vectors are expressed in the local coordinate frame [that is, all quantities in Eq.~\eqref{eq:second order detail} with tildes], the first three terms are spin preserving (connect states with the sum of spin projections along the local spin quantization axes), the next four terms change the sum by one (representing a single spin flip), and the last two terms induce double flips. The complex conjugates are defined as
\be
A^{m*}_\pm \equiv ({\bf r^\prime}_\pm \cdot {\bf A}^{m})^* = A^{m*}_x \mp {\rm i} A^{m*}_y =  {\bf r^\prime}_\mp \cdot ({\bf A}^{m})^*.
\ee

To find the effective Hamiltonian, it remains to evaluate the vectors ${\bf A}$. For electrons, we assume that the magnetic field dominates the total field for impurities. Equation \eqref{eq:aux} gives ($J=1/2$, $J^\prime=-1/2$)
\be
{\bf A}^n = -(\beta/2) |\Phi_G({\bf R}_n)|^2 R_{U_n} [{\bf r^\prime}_- ], 
\ee
where we remind that the vectors ${\bf \hat{x}^\prime}$, ${\bf \hat{y}^\prime}$, and ${\bf \hat{s}}_0={\bf \hat{z}^\prime}$ form an orthonormal set, with ${\bf \hat{s}}_0$ along the external field. Expanding the rotation operator in the lowest order in the spin-orbit length we finally find
\be
A^n_-\sim -(\beta/2) |\Phi_G({\bf R}_n)|^2, \quad A^n_+, A^n_z \sim O(r_n/l) A^n_-,
\ee
from where Eq.~\eqref{eq:spin breaking} of the main text follows.

For holes, we find the vectors ${\bf A}$ corresponding to the interaction mediated by the -3/2 state from Eqs.~\eqref{eq:ME for J 2} and Eq.~\eqref{eq:aux4} as
\be \begin{split}
{\bf r^\prime}_+ \cdot {\bf A} &= -\beta b^2 (a / |a|)(\cos\phi + 1),\\
{\bf r^\prime}_- \cdot {\bf A} &= -\sqrt{3}\beta \Phi b (a / |a|)(\cos\phi + 1),\\
{\bf \hat{z}^\prime} \cdot {\bf A} &= -\sqrt{3}\beta \Phi b (a / |a|)\cos\phi.
\end{split}
\ee
The largest spin-non-conserving term of the corresponding effective Hamiltonian is
\be
H^{nm}_{++} \sim \tilde{I}^n_+ \tilde{I}^m_+ \left( \frac{1}{E_G-E_p} 2\sqrt{3}(\beta \Phi^2)^2 \lambda_0^2 \right).
\label{eq:effective plusplus}
\ee

The vectors ${\bf A}$ for the spin alike light hole, using Eqs.~\eqref{eq:ME for J 3} follow as
\be \begin{split}
{\bf r^\prime}_+ \cdot {\bf A} &= -\sqrt{3}\beta \Phi^* \Phi^\prime (a^* / |a|)(\cos\phi + 1)/2,\\
{\bf r^\prime}_- \cdot {\bf A} &= -\beta \Phi^\prime b^* (\cos\phi + 1),\\
{\bf \hat{z}^\prime} \cdot {\bf A} &= -(\beta/2) (3a^\prime \Phi^* + a^*\Phi^\prime) \cos\phi.
\end{split}
\ee

From the above results, we see that the largest spin-non-conserving terms in the effective interaction are 
\be
H^{nm}_{++} \sim \tilde{I}^n_+ \tilde{I}^m_+ \left( \frac{1}{\Delta_{lh}} \sqrt{3} (\beta \Phi^2)^2\lambda_0 \right),
\ee
and
\be
H^{nm}_{+z} \sim \tilde{I}^n_+ \tilde{I}^m_z \left( \frac{1}{\Delta_{lh}} (3/4)\sqrt{3} (\beta \Phi^2)^2\lambda_1^\prime \right).
\label{eq:effective plusplus 2}
\ee
Similarly as before, a differentiation with respect to the position, which enters the impurity-phonon rates, brings in an additional factor $1/\delta$ in Eqs.~\eqref{eq:effective plusplus}-\eqref{eq:effective plusplus 2}.

\section{Materials parameters}

\label{app:parameters}

For the electronic case we assume a GaAs/AlGaAs heterostructure with the following parameters: electron effective mass $m=0.067\,m_0$, in-plane confinement length $l=30$ nm, quantum well width $w=8$ nm, spin-orbit length $l_{so} \sim$ 1 $\mu$m, electron-nuclear coupling $\beta=4\, \mu$eV nm$^3$, material density $\rho=5300$ kg/m$^3$, phonon velocities $c_l=5290$ m/s and $c_t=2480$ m/s, conduction band piezoelectric $\Xi=1.4\times 10^9$ eV/m and deformation $\sigma=10$ eV potentials. The g-factors and corresponding energy scales are given in Table \ref{tab:magnetic}.  

For the hole case, we list the Luttinger parameters $\gamma_1/\gamma_2/\gamma_3$ of GaAs:\cite{binggeli1991:PRB,shanabrook1989:PRB} 7.1/2/2.9, CdTe:\cite{friedrich1994:JPCM}4.1/1.1/1.6, and 
ZnTe:\cite{wagner1992:JCG} 3.8/0.7/1.3. We take ZnTe as the material of our choice, with $\rho=5650$ kg/m$^3$, $c_l=3550$ m/s, $c_t=2358$ m/s, $\sigma=5$ eV, $\Xi=3.4\times 10^8$ eV/m. We set the heavy hole orbital energy $\hbar \Omega_{hh}$ to 20 meV, which gives $l_{hh}=4.19$ nm, $l_{lh}=3.82$ nm and $\hbar \Omega_{lh} = 17$ meV. We set the light-heavy hole splitting $\Delta_{lh}$ to 100 meV, which gives $w=3.24$ nm. The hole-impurity interaction strength is $\beta = 1/3$ eV $a_0^3/4$, with $a_0=0.61$ nm the lattice constant. We assume the Mn impurities concentration is given as $x_{\rm Mn}$, the ratio of cation replaced by Mn atoms, typically of the order of 1\%.

Properly normalized, the ground state follows from Eq.~\eqref{eq:hole 1-st order states} as
\be
\Phi^J_{00,0}(r,\phi,z) = \sqrt{\frac{2}{w}} \sin\left( \frac{\pi z} {w} \right) \frac{1}{\sqrt{\pi} l_J} \exp \left( -\frac{r^2}{2l_J^2} \right),
\ee
from where the quantum dot volume estimate
\begin{equation}
V=1/\int {\rm d}^3{\bf R}\,|\Phi({\bf R})|^4 = (4\pi/3) w l_J^2,
\end{equation}
gives values in Table \ref{tab:volume}.
\begin{table}
\begin{tabular}{|l|ll|}
\hline
quantity & electron &hole\\
\hline
$V$ & 3 $\times10^4$ nm$^3$&238 nm$^3$\\
$N$ &  1.3 $\times\,10^6$ & $x_{\rm Mn} \times$ 5278\\
$\beta / V $ & 0.13 neV & 64 $\mu$eV\\
\hline
\end{tabular}
\caption{Effective volume, number of impurities and the coupling energy scale.}
\label{tab:volume}
\end{table}

\begin{table}
\begin{tabular}{|l|llll|}
\hline
quantity & nucleus&Mn&electron&hole\\
\hline
$g$ & 1.2 &2 &-0.44 & -2/3\\
$\mu$ & $\mu_N$ & $\mu_B$ &$\mu_B$ &$\mu_B$\\
$I$ or $J$ &3/2& 5/2 &1/2 &3/2\\
$g \mu I$ &  57 neV/T  & 290 $\mu$eV/T& 12.7 $\mu$eV/T &58 $\mu$eV/T\\
$g \mu I / k_B$ &  660 $\mu$K/T  & 3.4 K/T& 146 mK/T &670 mK/T\\
$B_{\rm eff}$  & 66 peV & 96 $\mu$eV &290  neV & 13.8 $x_{\rm Mn}^{1/2}$ meV\\
$B_{\rm eff}/(g \mu I)$ & 1.2 mT & 331 mT & 23 mT & 237 $x_{\rm Mn}^{1/2}$ T\\
\hline
\end{tabular}
\caption{g-factor, magnetic moment, spin, the corresponding energy scale and the effective field $B_{\rm eff}$. For GaAs nuclear spins, the g-factor is the average over naturally abundant isotopes. The effective field for impurities is the Knight field, $B_{\rm eff}=J (\beta / V)$, for the particle it is the Overhauser field, $B_{\rm eff}=\sqrt{NI(I+1)} (\beta / V)$.}
\label{tab:magnetic}
\end{table}

From the above parameters, the hole admixture coefficients follow as $\lambda_0 \approx 0.053$, $\lambda_1 \approx 0.050$, $\lambda_0^\prime \approx 0.11$, and $\lambda_1^\prime \approx 0.15$.

\section{Alternative hole mixing mechanisms}

\label{app:alternative mixing}

In this section we estimate two additional light-heavy hole admixture sources, namely the spin-orbit coupling to an external electric field (the Rashba spin-orbit interaction) and Overhauser impurity field itself. We quantify the resulting light-heavy hole admixture by calculating the corresponding coefficients $\lambda$. We find that these alternatives lead to negligible amount of admixture, compared to the terms of the Kohn-Luttinger Hamiltonian we considered in the main text.

In the presence of electric field ${\bf E}$ there appear the following term in the hole Hamiltonian (in the notation of Ref.~\onlinecite{winkler})
\be
H^{r}_{8v8v} = r^{8v8v}_{41}{\bf E} \cdot {\bf J} \times {\bf k}.
\ee
Assuming, for convenience, that the field is perpendicular to the heterostructure interface ${\bf E}={\bf \hat{z}} E$ this term translates into our notation of Sec.~\ref{app:KLH} as
\be
H^{so}_{\pm 3/2,\pm 1/2} = \pm r^{8v8v}_{41} E \sqrt{3} k_\mp /2,
\ee
where the material parameter $r^{8v8v}_{41}$ we estimate by its ZnSe value of $-4.1$ e\AA$^2$. Taking Eq.~\eqref{eq:hole 1-st order states} for the heavy hole ground state $|\Psi_{3/2}\rangle$, the Rashba spin-orbit interaction leads to an admixture of the light hole state $|1/2\rangle \otimes|\Phi^{lh}_{01,0}\rangle$ with a coefficient
\be
\lambda_{so} = -\frac{\rm i}{2} \frac{\sqrt{3} r^{8v8v}_{41} E }{\Delta_{lh}+\hbar \Omega_{lh}}.
\ee
To quantify its value, it remains put in the value for the electric field $E$. We estimate it to be of order $10^7$ V/m, using a very crude electrostatic model; namely we take the hole to be a point charge in the center of a half sphere of a radius $l \sim 5$ nm and the electron to be a uniform classical charge density $\sigma$ on the half-sphere, which corresponds to the type II-quantum dot populated by a single exciton. The resulting internal field $E=\sigma/2\epsilon$ is smaller than the piezoelectric field accompanying the phonon, Eq.~\eqref{eq:alpha eff 2}, which we estimated to be of order $10^8$ V/m. Inserting this latter value for $E$, we find $\lambda^{so} \approx \surd{3}\times 10^{-3}$, so that both the internal electric field as well as phonon induced fields induce light-heavy hole admixtures much smaller than those we considered in the main text. 

One may also consider the collective field of the impurities (the Overhauser field) as a source of the light-heavy hole mixing. Using Eqs.~\eqref{eq:ME for J 3} we estimate the arising admixture of the light hole state $|1/2\rangle \otimes|\Phi^{lh}_{00,0}\rangle$ into the heavy hole ground state
\be
|\lambda_{Mn}|^2 \approx \frac{(\beta/V)^2 N I(I+1)}{2\Delta_{lh}}.
\ee
The previous is a typical value, the phase being fixed by the microscopic state of the Mn ensemble. Evaluating for our parameters we get $\lambda_{Mn}\approx 7\times 10^{-3}$, typically an order of magnitude smaller admixture as that consider in the main text, so we can again neglect this admixture source.

\bibliography{../references/quantum_dot}

\begin{thebibliography}{91}
\expandafter\ifx\csname natexlab\endcsname\relax\def\natexlab#1{#1}\fi
\expandafter\ifx\csname bibnamefont\endcsname\relax
  \def\bibnamefont#1{#1}\fi
\expandafter\ifx\csname bibfnamefont\endcsname\relax
  \def\bibfnamefont#1{#1}\fi
\expandafter\ifx\csname citenamefont\endcsname\relax
  \def\citenamefont#1{#1}\fi
\expandafter\ifx\csname url\endcsname\relax
  \def\url#1{\texttt{#1}}\fi
\expandafter\ifx\csname urlprefix\endcsname\relax\def\urlprefix{URL }\fi
\providecommand{\bibinfo}[2]{#2}
\providecommand{\eprint}[2][]{\url{#2}}

\bibitem[{\citenamefont{Loss and DiVincenzo}(1998)}]{loss1998:PRA}
\bibinfo{author}{\bibfnamefont{D.}~\bibnamefont{Loss}} \bibnamefont{and}
  \bibinfo{author}{\bibfnamefont{D.~P.} \bibnamefont{DiVincenzo}},
  \bibinfo{journal}{Phys. Rev. A} \textbf{\bibinfo{volume}{57}},
  \bibinfo{pages}{120} (\bibinfo{year}{1998}).

\bibitem[{\citenamefont{Hanson et~al.}(2007)\citenamefont{Hanson, Kouwenhoven,
  Petta, Tarucha, and Vandersypen}}]{hanson2007:RMP}
\bibinfo{author}{\bibfnamefont{R.}~\bibnamefont{Hanson}},
  \bibinfo{author}{\bibfnamefont{L.~P.} \bibnamefont{Kouwenhoven}},
  \bibinfo{author}{\bibfnamefont{J.~R.} \bibnamefont{Petta}},
  \bibinfo{author}{\bibfnamefont{S.}~\bibnamefont{Tarucha}}, \bibnamefont{and}
  \bibinfo{author}{\bibfnamefont{L.~M.~K.} \bibnamefont{Vandersypen}},
  \bibinfo{journal}{Rev. Mod. Phys.} \textbf{\bibinfo{volume}{79}},
  \bibinfo{pages}{1217} (\bibinfo{year}{2007}).

\bibitem[{\citenamefont{Taylor et~al.}(2007)\citenamefont{Taylor, Petta,
  Johnson, Yacoby, Marcus, and Lukin}}]{taylor2007:PRB}
\bibinfo{author}{\bibfnamefont{J.~M.} \bibnamefont{Taylor}},
  \bibinfo{author}{\bibfnamefont{J.~R.} \bibnamefont{Petta}},
  \bibinfo{author}{\bibfnamefont{A.~C.} \bibnamefont{Johnson}},
  \bibinfo{author}{\bibfnamefont{A.}~\bibnamefont{Yacoby}},
  \bibinfo{author}{\bibfnamefont{C.~M.} \bibnamefont{Marcus}},
  \bibnamefont{and} \bibinfo{author}{\bibfnamefont{M.~D.} \bibnamefont{Lukin}},
  \bibinfo{journal}{Phys. Rev. B} \textbf{\bibinfo{volume}{76}},
  \bibinfo{pages}{035315} (\bibinfo{year}{2007}).

\bibitem[{\citenamefont{Koppens et~al.}(2006)\citenamefont{Koppens, Buizert,
  Tielrooij, Vink, Nowack, Meunier, Kouwenhoven, and
  Vandersypen}}]{koppens2006:N}
\bibinfo{author}{\bibfnamefont{F.~H.~L.} \bibnamefont{Koppens}},
  \bibinfo{author}{\bibfnamefont{C.}~\bibnamefont{Buizert}},
  \bibinfo{author}{\bibfnamefont{K.~J.} \bibnamefont{Tielrooij}},
  \bibinfo{author}{\bibfnamefont{I.~T.} \bibnamefont{Vink}},
  \bibinfo{author}{\bibfnamefont{K.~C.} \bibnamefont{Nowack}},
  \bibinfo{author}{\bibfnamefont{T.}~\bibnamefont{Meunier}},
  \bibinfo{author}{\bibfnamefont{L.~P.} \bibnamefont{Kouwenhoven}},
  \bibnamefont{and} \bibinfo{author}{\bibfnamefont{L.~M.~K.}
  \bibnamefont{Vandersypen}}, \bibinfo{journal}{Nature}
  \textbf{\bibinfo{volume}{442}}, \bibinfo{pages}{766} (\bibinfo{year}{2006}).

\bibitem[{\citenamefont{Nowack et~al.}(2007)\citenamefont{Nowack, Koppens,
  Nazarov, and Vandersypen}}]{nowack2007:S}
\bibinfo{author}{\bibfnamefont{K.~C.} \bibnamefont{Nowack}},
  \bibinfo{author}{\bibfnamefont{F.~H.~L.} \bibnamefont{Koppens}},
  \bibinfo{author}{\bibfnamefont{Y.~V.} \bibnamefont{Nazarov}},
  \bibnamefont{and} \bibinfo{author}{\bibfnamefont{L.~M.~K.}
  \bibnamefont{Vandersypen}}, \bibinfo{journal}{Science}
  \textbf{\bibinfo{volume}{318}}, \bibinfo{pages}{1430} (\bibinfo{year}{2007}).

\bibitem[{\citenamefont{Obata et~al.}(2010)\citenamefont{Obata, Pioro-Ladriere,
  Tokura, Shin, Kubo, Yoshida, Taniyama, and Tarucha}}]{obata2010:PRB}
\bibinfo{author}{\bibfnamefont{T.}~\bibnamefont{Obata}},
  \bibinfo{author}{\bibfnamefont{M.}~\bibnamefont{Pioro-Ladriere}},
  \bibinfo{author}{\bibfnamefont{Y.}~\bibnamefont{Tokura}},
  \bibinfo{author}{\bibfnamefont{Y.-S.} \bibnamefont{Shin}},
  \bibinfo{author}{\bibfnamefont{T.}~\bibnamefont{Kubo}},
  \bibinfo{author}{\bibfnamefont{K.}~\bibnamefont{Yoshida}},
  \bibinfo{author}{\bibfnamefont{T.}~\bibnamefont{Taniyama}}, \bibnamefont{and}
  \bibinfo{author}{\bibfnamefont{S.}~\bibnamefont{Tarucha}},
  \bibinfo{journal}{Phys Rev B} \textbf{\bibinfo{volume}{81}},
  \bibinfo{pages}{085317} (\bibinfo{year}{2010}).

\bibitem[{\citenamefont{Press et~al.}(2008)\citenamefont{Press, Ladd, Zhang,
  and Yamamoto}}]{press2008:N}
\bibinfo{author}{\bibfnamefont{D.}~\bibnamefont{Press}},
  \bibinfo{author}{\bibfnamefont{T.~D.} \bibnamefont{Ladd}},
  \bibinfo{author}{\bibfnamefont{B.}~\bibnamefont{Zhang}}, \bibnamefont{and}
  \bibinfo{author}{\bibfnamefont{Y.}~\bibnamefont{Yamamoto}},
  \bibinfo{journal}{Nature} \textbf{\bibinfo{volume}{456}},
  \bibinfo{pages}{218} (\bibinfo{year}{2008}).

\bibitem[{\citenamefont{Fabian et~al.}(2007)\citenamefont{Fabian,
  {Matos-Abiague}, Ertler, Stano, and {\v{Z}uti\'{c}}}}]{fabian2007:APS}
\bibinfo{author}{\bibfnamefont{J.}~\bibnamefont{Fabian}},
  \bibinfo{author}{\bibfnamefont{A.}~\bibnamefont{{Matos-Abiague}}},
  \bibinfo{author}{\bibfnamefont{C.}~\bibnamefont{Ertler}},
  \bibinfo{author}{\bibfnamefont{P.}~\bibnamefont{Stano}}, \bibnamefont{and}
  \bibinfo{author}{\bibfnamefont{I.}~\bibnamefont{{\v{Z}uti\'{c}}}},
  \bibinfo{journal}{Acta Phys. Slov.} \textbf{\bibinfo{volume}{57}},
  \bibinfo{pages}{565} (\bibinfo{year}{2007}), \bibinfo{note}{arXiv:0711.1461}.

\bibitem[{\citenamefont{{\v{Z}uti\'c} et~al.}(2004)\citenamefont{{\v{Z}uti\'c},
  Fabian, and {Das Sarma}}}]{zutic2004:RMP}
\bibinfo{author}{\bibfnamefont{I.}~\bibnamefont{{\v{Z}uti\'c}}},
  \bibinfo{author}{\bibfnamefont{J.}~\bibnamefont{Fabian}}, \bibnamefont{and}
  \bibinfo{author}{\bibfnamefont{S.}~\bibnamefont{{Das Sarma}}},
  \bibinfo{journal}{Rev. Mod. Phys.} \textbf{\bibinfo{volume}{76}},
  \bibinfo{pages}{323} (\bibinfo{year}{2004}).

\bibitem[{\citenamefont{Coish and Baugh}(2009)}]{coish2009:PSS}
\bibinfo{author}{\bibfnamefont{W.~A.} \bibnamefont{Coish}} \bibnamefont{and}
  \bibinfo{author}{\bibfnamefont{J.}~\bibnamefont{Baugh}},
  \bibinfo{journal}{Phys. Stat. Sol. B} \textbf{\bibinfo{volume}{246}},
  \bibinfo{pages}{2203} (\bibinfo{year}{2009}).

\bibitem[{\citenamefont{Erlingsson et~al.}(2001)\citenamefont{Erlingsson,
  Nazarov, and {Fa\v{l}ko}}}]{erlingsson2001:PRB}
\bibinfo{author}{\bibfnamefont{S.~I.} \bibnamefont{Erlingsson}},
  \bibinfo{author}{\bibfnamefont{Y.~V.} \bibnamefont{Nazarov}},
  \bibnamefont{and} \bibinfo{author}{\bibfnamefont{V.~I.}
  \bibnamefont{{Fa\v{l}ko}}}, \bibinfo{journal}{Phys. Rev. B}
  \textbf{\bibinfo{volume}{64}}, \bibinfo{pages}{195306}
  (\bibinfo{year}{2001}).

\bibitem[{\citenamefont{Khaetskii et~al.}(2002)\citenamefont{Khaetskii, Loss,
  and Glazman}}]{khaetskii2002:PRL}
\bibinfo{author}{\bibfnamefont{A.~V.} \bibnamefont{Khaetskii}},
  \bibinfo{author}{\bibfnamefont{D.}~\bibnamefont{Loss}}, \bibnamefont{and}
  \bibinfo{author}{\bibfnamefont{L.}~\bibnamefont{Glazman}},
  \bibinfo{journal}{Phys. Rev. Lett.} \textbf{\bibinfo{volume}{88}},
  \bibinfo{pages}{186802} (\bibinfo{year}{2002}).

\bibitem[{\citenamefont{Merkulov et~al.}(2002)\citenamefont{Merkulov, Efros,
  and Rosen}}]{merkulov2002:PRB}
\bibinfo{author}{\bibfnamefont{I.~A.} \bibnamefont{Merkulov}},
  \bibinfo{author}{\bibfnamefont{A.~L.} \bibnamefont{Efros}}, \bibnamefont{and}
  \bibinfo{author}{\bibfnamefont{M.}~\bibnamefont{Rosen}},
  \bibinfo{journal}{Phys. Rev. B} \textbf{\bibinfo{volume}{65}},
  \bibinfo{pages}{205309} (\bibinfo{year}{2002}).

\bibitem[{\citenamefont{Coish et~al.}(2006)\citenamefont{Coish, Golovach, and
  Egues}}]{coish2006:PSS}
\bibinfo{author}{\bibfnamefont{W.~A.} \bibnamefont{Coish}},
  \bibinfo{author}{\bibfnamefont{V.~N.} \bibnamefont{Golovach}},
  \bibnamefont{and} \bibinfo{author}{\bibfnamefont{J.~C.} \bibnamefont{Egues}},
  \bibinfo{journal}{Phys. Stat. Sol. B} \textbf{\bibinfo{volume}{243}},
  \bibinfo{pages}{3658} (\bibinfo{year}{2006}).

\bibitem[{\citenamefont{Cywinski
  et~al.}(2009{\natexlab{a}})\citenamefont{Cywinski, Witzel, and {Das
  Sarma}}}]{cywinski2009:PRB}
\bibinfo{author}{\bibfnamefont{L.}~\bibnamefont{Cywinski}},
  \bibinfo{author}{\bibfnamefont{W.~M.} \bibnamefont{Witzel}},
  \bibnamefont{and} \bibinfo{author}{\bibfnamefont{S.}~\bibnamefont{{Das
  Sarma}}}, \bibinfo{journal}{Phys. Rev. B} \textbf{\bibinfo{volume}{79}},
  \bibinfo{pages}{245314} (\bibinfo{year}{2009}{\natexlab{a}}).

\bibitem[{\citenamefont{Cywinski
  et~al.}(2009{\natexlab{b}})\citenamefont{Cywinski, Witzel, and {Das
  Sarma}}}]{cywinski2009:PRL}
\bibinfo{author}{\bibfnamefont{L.}~\bibnamefont{Cywinski}},
  \bibinfo{author}{\bibfnamefont{W.~M.} \bibnamefont{Witzel}},
  \bibnamefont{and} \bibinfo{author}{\bibfnamefont{S.}~\bibnamefont{{Das
  Sarma}}}, \bibinfo{journal}{Phys. Rev. Lett.} \textbf{\bibinfo{volume}{102}},
  \bibinfo{pages}{057601} (\bibinfo{year}{2009}{\natexlab{b}}).

\bibitem[{\citenamefont{Merkulov et~al.}(2010)\citenamefont{Merkulov, Alvarez,
  Yakovlev, and Schulthess}}]{merkulov2010:PRB}
\bibinfo{author}{\bibfnamefont{I.~A.} \bibnamefont{Merkulov}},
  \bibinfo{author}{\bibfnamefont{G.}~\bibnamefont{Alvarez}},
  \bibinfo{author}{\bibfnamefont{D.~R.} \bibnamefont{Yakovlev}},
  \bibnamefont{and} \bibinfo{author}{\bibfnamefont{T.~C.}
  \bibnamefont{Schulthess}}, \bibinfo{journal}{Phys. Rev. B}
  \textbf{\bibinfo{volume}{81}}, \bibinfo{pages}{115107}
  (\bibinfo{year}{2010}).

\bibitem[{\citenamefont{Petta et~al.}(2005)\citenamefont{Petta, Johnson,
  Taylor, Laird, Yacoby, Lukin, Marcus, Hanson, and Gossard}}]{petta2005:S}
\bibinfo{author}{\bibfnamefont{J.~R.} \bibnamefont{Petta}},
  \bibinfo{author}{\bibfnamefont{A.~C.} \bibnamefont{Johnson}},
  \bibinfo{author}{\bibfnamefont{J.~M.} \bibnamefont{Taylor}},
  \bibinfo{author}{\bibfnamefont{E.~A.} \bibnamefont{Laird}},
  \bibinfo{author}{\bibfnamefont{A.}~\bibnamefont{Yacoby}},
  \bibinfo{author}{\bibfnamefont{M.~D.} \bibnamefont{Lukin}},
  \bibinfo{author}{\bibfnamefont{C.~M.} \bibnamefont{Marcus}},
  \bibinfo{author}{\bibfnamefont{M.~P.} \bibnamefont{Hanson}},
  \bibnamefont{and} \bibinfo{author}{\bibfnamefont{A.~C.}
  \bibnamefont{Gossard}}, \bibinfo{journal}{Science}
  \textbf{\bibinfo{volume}{309}}, \bibinfo{pages}{2180} (\bibinfo{year}{2005}).

\bibitem[{\citenamefont{Koppens et~al.}(2008)\citenamefont{Koppens, Nowack, and
  Vandersypen}}]{koppens2008:PRL}
\bibinfo{author}{\bibfnamefont{F.~H.~L.} \bibnamefont{Koppens}},
  \bibinfo{author}{\bibfnamefont{K.~C.} \bibnamefont{Nowack}},
  \bibnamefont{and} \bibinfo{author}{\bibfnamefont{L.~M.~K.}
  \bibnamefont{Vandersypen}}, \bibinfo{journal}{Phys. Rev. Lett.}
  \textbf{\bibinfo{volume}{100}}, \bibinfo{pages}{236802}
  (\bibinfo{year}{2008}).

\bibitem[{\citenamefont{Bluhm et~al.}(2010)\citenamefont{Bluhm, Foletti, Neder,
  Rudner, Mahalu, Umansky, and Yacoby}}]{bluhm2010:N}
\bibinfo{author}{\bibfnamefont{H.}~\bibnamefont{Bluhm}},
  \bibinfo{author}{\bibfnamefont{S.}~\bibnamefont{Foletti}},
  \bibinfo{author}{\bibfnamefont{I.}~\bibnamefont{Neder}},
  \bibinfo{author}{\bibfnamefont{M.}~\bibnamefont{Rudner}},
  \bibinfo{author}{\bibfnamefont{D.}~\bibnamefont{Mahalu}},
  \bibinfo{author}{\bibfnamefont{V.}~\bibnamefont{Umansky}}, \bibnamefont{and}
  \bibinfo{author}{\bibfnamefont{A.}~\bibnamefont{Yacoby}},
  \bibinfo{journal}{Nature Phys.} \textbf{\bibinfo{volume}{7}},
  \bibinfo{pages}{109} (\bibinfo{year}{2010}).

\bibitem[{\citenamefont{Schliemann
  et~al.}(2003{\natexlab{a}})\citenamefont{Schliemann, Khaetskii, and
  Loss}}]{schliemann2003:JPCM}
\bibinfo{author}{\bibfnamefont{J.}~\bibnamefont{Schliemann}},
  \bibinfo{author}{\bibfnamefont{A.}~\bibnamefont{Khaetskii}},
  \bibnamefont{and} \bibinfo{author}{\bibfnamefont{D.}~\bibnamefont{Loss}},
  \bibinfo{journal}{J. Phys: Condens. Matter} \textbf{\bibinfo{volume}{15}},
  \bibinfo{pages}{R1809} (\bibinfo{year}{2003}{\natexlab{a}}).

\bibitem[{\citenamefont{Reilly et~al.}(2008)\citenamefont{Reilly, Taylor,
  Petta, Marcus, Hanson, and Gossard}}]{reilly2008:S}
\bibinfo{author}{\bibfnamefont{D.~J.} \bibnamefont{Reilly}},
  \bibinfo{author}{\bibfnamefont{J.~M.} \bibnamefont{Taylor}},
  \bibinfo{author}{\bibfnamefont{J.~R.} \bibnamefont{Petta}},
  \bibinfo{author}{\bibfnamefont{C.~M.} \bibnamefont{Marcus}},
  \bibinfo{author}{\bibfnamefont{M.~P.} \bibnamefont{Hanson}},
  \bibnamefont{and} \bibinfo{author}{\bibfnamefont{A.~C.}
  \bibnamefont{Gossard}}, \bibinfo{journal}{Science}
  \textbf{\bibinfo{volume}{321}}, \bibinfo{pages}{5890} (\bibinfo{year}{2008}).

\bibitem[{\citenamefont{Bracker et~al.}(2005)\citenamefont{Bracker, Stinaff,
  Gammon, Ware, Tischler, Shabaev, Efros, Park, Gershoni, Korenev
  et~al.}}]{bracker2005:PRL}
\bibinfo{author}{\bibfnamefont{A.~S.} \bibnamefont{Bracker}},
  \bibinfo{author}{\bibfnamefont{E.~A.} \bibnamefont{Stinaff}},
  \bibinfo{author}{\bibfnamefont{D.}~\bibnamefont{Gammon}},
  \bibinfo{author}{\bibfnamefont{M.~E.} \bibnamefont{Ware}},
  \bibinfo{author}{\bibfnamefont{J.~G.} \bibnamefont{Tischler}},
  \bibinfo{author}{\bibfnamefont{A.}~\bibnamefont{Shabaev}},
  \bibinfo{author}{\bibfnamefont{A.~L.} \bibnamefont{Efros}},
  \bibinfo{author}{\bibfnamefont{D.}~\bibnamefont{Park}},
  \bibinfo{author}{\bibfnamefont{D.}~\bibnamefont{Gershoni}},
  \bibinfo{author}{\bibfnamefont{V.~L.} \bibnamefont{Korenev}},
  \bibnamefont{et~al.}, \bibinfo{journal}{Phys. Rev. Lett.}
  \textbf{\bibinfo{volume}{94}}, \bibinfo{pages}{047402}
  (\bibinfo{year}{2005}).

\bibitem[{\citenamefont{Lai et~al.}(2006)\citenamefont{Lai, Maletinsky,
  Badolato, and Imamoglu}}]{lai2006:PRL}
\bibinfo{author}{\bibfnamefont{C.~W.} \bibnamefont{Lai}},
  \bibinfo{author}{\bibfnamefont{P.}~\bibnamefont{Maletinsky}},
  \bibinfo{author}{\bibfnamefont{A.}~\bibnamefont{Badolato}}, \bibnamefont{and}
  \bibinfo{author}{\bibfnamefont{A.}~\bibnamefont{Imamoglu}},
  \bibinfo{journal}{Phys. Rev. Lett.} \textbf{\bibinfo{volume}{96}},
  \bibinfo{pages}{167403} (\bibinfo{year}{2006}).

\bibitem[{\citenamefont{Högele et~al.}(unpublished)\citenamefont{Högele,
  Kroner, Latta, Claassen, Carusotto, Bulutay, and Imamoglu}}]{hogele2011:CM}
\bibinfo{author}{\bibfnamefont{A.}~\bibnamefont{Högele}},
  \bibinfo{author}{\bibfnamefont{M.}~\bibnamefont{Kroner}},
  \bibinfo{author}{\bibfnamefont{C.}~\bibnamefont{Latta}},
  \bibinfo{author}{\bibfnamefont{M.}~\bibnamefont{Claassen}},
  \bibinfo{author}{\bibfnamefont{I.}~\bibnamefont{Carusotto}},
  \bibinfo{author}{\bibfnamefont{C.}~\bibnamefont{Bulutay}}, \bibnamefont{and}
  \bibinfo{author}{\bibfnamefont{A.}~\bibnamefont{Imamoglu}},
  \bibinfo{journal}{arxiv:1110.5524}  (\bibinfo{year}{unpublished}).

\bibitem[{\citenamefont{Seufert et~al.}(2001)\citenamefont{Seufert, Bacher,
  Scheibner, Forchel, Lee, Dobrowolska, and Furdyna}}]{seufert2001:PRL}
\bibinfo{author}{\bibfnamefont{J.}~\bibnamefont{Seufert}},
  \bibinfo{author}{\bibfnamefont{G.}~\bibnamefont{Bacher}},
  \bibinfo{author}{\bibfnamefont{M.}~\bibnamefont{Scheibner}},
  \bibinfo{author}{\bibfnamefont{A.}~\bibnamefont{Forchel}},
  \bibinfo{author}{\bibfnamefont{S.}~\bibnamefont{Lee}},
  \bibinfo{author}{\bibfnamefont{M.}~\bibnamefont{Dobrowolska}},
  \bibnamefont{and} \bibinfo{author}{\bibfnamefont{J.~K.}
  \bibnamefont{Furdyna}}, \bibinfo{journal}{Phys. Rev. Lett.}
  \textbf{\bibinfo{volume}{88}}, \bibinfo{pages}{027402}
  (\bibinfo{year}{2001}).

\bibitem[{\citenamefont{Besombes et~al.}(2004)\citenamefont{Besombes, L\'eger,
  Maingault, Ferrand, Mariette, and Cibert}}]{besombes2004:PRL}
\bibinfo{author}{\bibfnamefont{L.}~\bibnamefont{Besombes}},
  \bibinfo{author}{\bibfnamefont{Y.}~\bibnamefont{L\'eger}},
  \bibinfo{author}{\bibfnamefont{L.}~\bibnamefont{Maingault}},
  \bibinfo{author}{\bibfnamefont{D.}~\bibnamefont{Ferrand}},
  \bibinfo{author}{\bibfnamefont{H.}~\bibnamefont{Mariette}}, \bibnamefont{and}
  \bibinfo{author}{\bibfnamefont{J.}~\bibnamefont{Cibert}},
  \bibinfo{journal}{Phys. Rev. Lett.} \textbf{\bibinfo{volume}{93}},
  \bibinfo{pages}{207403} (\bibinfo{year}{2004}).

\bibitem[{\citenamefont{Xiu et~al.}(2010)\citenamefont{Xiu, Wang, Kim, Zhou,
  Kou, Han, Kawakami, Zou, and Wang}}]{xiu2010:ACSN}
\bibinfo{author}{\bibfnamefont{F.}~\bibnamefont{Xiu}},
  \bibinfo{author}{\bibfnamefont{Y.}~\bibnamefont{Wang}},
  \bibinfo{author}{\bibfnamefont{J.}~\bibnamefont{Kim}},
  \bibinfo{author}{\bibfnamefont{Y.}~\bibnamefont{Zhou}},
  \bibinfo{author}{\bibfnamefont{X.}~\bibnamefont{Kou}},
  \bibinfo{author}{\bibfnamefont{W.}~\bibnamefont{Han}},
  \bibinfo{author}{\bibfnamefont{R.~K.} \bibnamefont{Kawakami}},
  \bibinfo{author}{\bibfnamefont{J.}~\bibnamefont{Zou}}, \bibnamefont{and}
  \bibinfo{author}{\bibfnamefont{K.~L.} \bibnamefont{Wang}},
  \bibinfo{journal}{ACS Nano.} \textbf{\bibinfo{volume}{4}},
  \bibinfo{pages}{4948} (\bibinfo{year}{2010}).

\bibitem[{\citenamefont{Klopotowski et~al.}(2011)\citenamefont{Klopotowski,
  Cywi\'nski, Wojnar, Voliotis, Fronc, Kazimierczuk, Golnik, Ravaro, Grousson,
  Karczewski et~al.}}]{klopotowski2011:PRB}
\bibinfo{author}{\bibfnamefont{L.}~\bibnamefont{Klopotowski}},
  \bibinfo{author}{\bibfnamefont{L.}~\bibnamefont{Cywi\'nski}},
  \bibinfo{author}{\bibfnamefont{P.}~\bibnamefont{Wojnar}},
  \bibinfo{author}{\bibfnamefont{V.}~\bibnamefont{Voliotis}},
  \bibinfo{author}{\bibfnamefont{K.}~\bibnamefont{Fronc}},
  \bibinfo{author}{\bibfnamefont{T.}~\bibnamefont{Kazimierczuk}},
  \bibinfo{author}{\bibfnamefont{A.}~\bibnamefont{Golnik}},
  \bibinfo{author}{\bibfnamefont{M.}~\bibnamefont{Ravaro}},
  \bibinfo{author}{\bibfnamefont{R.}~\bibnamefont{Grousson}},
  \bibinfo{author}{\bibfnamefont{G.}~\bibnamefont{Karczewski}},
  \bibnamefont{et~al.}, \bibinfo{journal}{Phys. Rev. B}
  \textbf{\bibinfo{volume}{83}}, \bibinfo{pages}{081306(R)}
  (\bibinfo{year}{2011}).

\bibitem[{\citenamefont{Maksimov et~al.}(2000)\citenamefont{Maksimov, Bacher,
  McDonald, Kulakovskii, Forchel, Becker, Landwehr, and
  Molenkamp}}]{maksimov2000:PRB}
\bibinfo{author}{\bibfnamefont{A.~A.} \bibnamefont{Maksimov}},
  \bibinfo{author}{\bibfnamefont{G.}~\bibnamefont{Bacher}},
  \bibinfo{author}{\bibfnamefont{A.}~\bibnamefont{McDonald}},
  \bibinfo{author}{\bibfnamefont{V.~D.} \bibnamefont{Kulakovskii}},
  \bibinfo{author}{\bibfnamefont{A.}~\bibnamefont{Forchel}},
  \bibinfo{author}{\bibfnamefont{C.~R.} \bibnamefont{Becker}},
  \bibinfo{author}{\bibfnamefont{G.}~\bibnamefont{Landwehr}}, \bibnamefont{and}
  \bibinfo{author}{\bibfnamefont{L.~W.} \bibnamefont{Molenkamp}},
  \bibinfo{journal}{Phys. Rev. B} \textbf{\bibinfo{volume}{62}},
  \bibinfo{pages}{R7767} (\bibinfo{year}{2000}).

\bibitem[{\citenamefont{Beaulac et~al.}(2009)\citenamefont{Beaulac, Schneider,
  Archer, Bacher, and Gamelin}}]{beaulac2009:S}
\bibinfo{author}{\bibfnamefont{R.}~\bibnamefont{Beaulac}},
  \bibinfo{author}{\bibfnamefont{L.}~\bibnamefont{Schneider}},
  \bibinfo{author}{\bibfnamefont{P.~I.} \bibnamefont{Archer}},
  \bibinfo{author}{\bibfnamefont{G.}~\bibnamefont{Bacher}}, \bibnamefont{and}
  \bibinfo{author}{\bibfnamefont{D.~R.} \bibnamefont{Gamelin}},
  \bibinfo{journal}{Science} \textbf{\bibinfo{volume}{325}},
  \bibinfo{pages}{973} (\bibinfo{year}{2009}).

\bibitem[{\citenamefont{Sellers et~al.}(2010)\citenamefont{Sellers,
  Oszwaldowski, Whiteside, Eginligil, Petrou, \v{Z}uti\'{c}, Chou, Fan,
  G.Petukhov, Kim et~al.}}]{sellers2010:PRB}
\bibinfo{author}{\bibfnamefont{I.~R.} \bibnamefont{Sellers}},
  \bibinfo{author}{\bibfnamefont{R.}~\bibnamefont{Oszwaldowski}},
  \bibinfo{author}{\bibfnamefont{V.~R.} \bibnamefont{Whiteside}},
  \bibinfo{author}{\bibfnamefont{M.}~\bibnamefont{Eginligil}},
  \bibinfo{author}{\bibfnamefont{A.}~\bibnamefont{Petrou}},
  \bibinfo{author}{\bibfnamefont{I.}~\bibnamefont{\v{Z}uti\'{c}}},
  \bibinfo{author}{\bibfnamefont{W.-C.} \bibnamefont{Chou}},
  \bibinfo{author}{\bibfnamefont{W.~C.} \bibnamefont{Fan}},
  \bibinfo{author}{\bibfnamefont{A.}~\bibnamefont{G.Petukhov}},
  \bibinfo{author}{\bibfnamefont{S.~J.} \bibnamefont{Kim}},
  \bibnamefont{et~al.}, \bibinfo{journal}{Phys. Rev. B}
  \textbf{\bibinfo{volume}{82}}, \bibinfo{pages}{195320}
  (\bibinfo{year}{2010}).

\bibitem[{\citenamefont{Govorov}(2005)}]{govorov2005:PRB}
\bibinfo{author}{\bibfnamefont{A.~O.} \bibnamefont{Govorov}},
  \bibinfo{journal}{Phys. Rev. B} \textbf{\bibinfo{volume}{72}},
  \bibinfo{pages}{075359} (\bibinfo{year}{2005}).

\bibitem[{\citenamefont{Govorov}(2008)}]{govorov2008:CRP}
\bibinfo{author}{\bibfnamefont{A.~O.} \bibnamefont{Govorov}},
  \bibinfo{journal}{C.R. Physique} \textbf{\bibinfo{volume}{9}},
  \bibinfo{pages}{857} (\bibinfo{year}{2008}).

\bibitem[{\citenamefont{Fern\'andez-Rossier and
  Brey}(2004)}]{fernandez-rossier2004:PRL}
\bibinfo{author}{\bibfnamefont{J.}~\bibnamefont{Fern\'andez-Rossier}}
  \bibnamefont{and} \bibinfo{author}{\bibfnamefont{L.}~\bibnamefont{Brey}},
  \bibinfo{journal}{Phys. rev. Lett.} \textbf{\bibinfo{volume}{93}},
  \bibinfo{pages}{117201} (\bibinfo{year}{2004}).

\bibitem[{\citenamefont{Qu and Hawrylak}(2006)}]{qu2006:PRL}
\bibinfo{author}{\bibfnamefont{F.}~\bibnamefont{Qu}} \bibnamefont{and}
  \bibinfo{author}{\bibfnamefont{P.}~\bibnamefont{Hawrylak}},
  \bibinfo{journal}{Phys. Rev. Lett.} \textbf{\bibinfo{volume}{96}},
  \bibinfo{pages}{157201} (\bibinfo{year}{2006}).

\bibitem[{\citenamefont{Nguyen and Peeters}(2008)}]{nguyen2008:PRB}
\bibinfo{author}{\bibfnamefont{N.~T.~T.} \bibnamefont{Nguyen}}
  \bibnamefont{and} \bibinfo{author}{\bibfnamefont{F.~M.}
  \bibnamefont{Peeters}}, \bibinfo{journal}{Phys. Rev. B}
  \textbf{\bibinfo{volume}{78}}, \bibinfo{pages}{045321}
  (\bibinfo{year}{2008}).

\bibitem[{\citenamefont{Abolfath et~al.}(2007)\citenamefont{Abolfath, Hawrylak,
  and {\v{Z}uti\'c}}}]{abolfath2007:PRL}
\bibinfo{author}{\bibfnamefont{R.~M.} \bibnamefont{Abolfath}},
  \bibinfo{author}{\bibfnamefont{P.}~\bibnamefont{Hawrylak}}, \bibnamefont{and}
  \bibinfo{author}{\bibfnamefont{I.}~\bibnamefont{{\v{Z}uti\'c}}},
  \bibinfo{journal}{Phys. Rev. Lett.} \textbf{\bibinfo{volume}{98}},
  \bibinfo{pages}{207203} (\bibinfo{year}{2007}).

\bibitem[{\citenamefont{Oszwaldowski et~al.}(2011)\citenamefont{Oszwaldowski,
  {\v{Z}uti\'c}, and Petukhov}}]{oswaldowski2011:PRL}
\bibinfo{author}{\bibfnamefont{R.}~\bibnamefont{Oszwaldowski}},
  \bibinfo{author}{\bibfnamefont{I.}~\bibnamefont{{\v{Z}uti\'c}}},
  \bibnamefont{and} \bibinfo{author}{\bibfnamefont{A.~G.}
  \bibnamefont{Petukhov}}, \bibinfo{journal}{Phys. Rev. Lett.}
  \textbf{\bibinfo{volume}{106}}, \bibinfo{pages}{177201}
  (\bibinfo{year}{2011}).

\bibitem[{\citenamefont{Lebedeva et~al.}(2010)\citenamefont{Lebedeva, Varpula,
  Novikov, and Kuivalainen}}]{lebedeva2010:PRB}
\bibinfo{author}{\bibfnamefont{N.}~\bibnamefont{Lebedeva}},
  \bibinfo{author}{\bibfnamefont{A.}~\bibnamefont{Varpula}},
  \bibinfo{author}{\bibfnamefont{S.}~\bibnamefont{Novikov}}, \bibnamefont{and}
  \bibinfo{author}{\bibfnamefont{P.}~\bibnamefont{Kuivalainen}},
  \bibinfo{journal}{Phys. Rev. B} \textbf{\bibinfo{volume}{81}},
  \bibinfo{pages}{235307} (\bibinfo{year}{2010}).

\bibitem[{\citenamefont{Yakovlev and Ossau}(2010)}]{yakovlev}
\bibinfo{author}{\bibfnamefont{D.~R.} \bibnamefont{Yakovlev}} \bibnamefont{and}
  \bibinfo{author}{\bibfnamefont{W.}~\bibnamefont{Ossau}}
  (\bibinfo{publisher}{Introduction to the Physics of Diluted Magnetic
  Semiconductors edited by J. Kossut and J. A. Gaj, Springer, Berlin},
  \bibinfo{year}{2010}).

\bibitem[{\citenamefont{Dietl and Spalek}(1983)}]{dietl1983:PRB}
\bibinfo{author}{\bibfnamefont{T.}~\bibnamefont{Dietl}} \bibnamefont{and}
  \bibinfo{author}{\bibfnamefont{J.}~\bibnamefont{Spalek}},
  \bibinfo{journal}{Phys. Rev. B} \textbf{\bibinfo{volume}{28}},
  \bibinfo{pages}{1548} (\bibinfo{year}{1983}).

\bibitem[{\citenamefont{Wolff}(1988)}]{wolff}
\bibinfo{author}{\bibfnamefont{P.~A.} \bibnamefont{Wolff}}
  (\bibinfo{publisher}{Semiconductors and Semimetals edited by J. K. Furdyna
  and J. Kossut (Academic Press, San Diego), Vol. 25}, \bibinfo{year}{1988}).

\bibitem[{\citenamefont{Nagaev}(1983)}]{nagaev}
\bibinfo{author}{\bibfnamefont{E.~L.} \bibnamefont{Nagaev}}
  (\bibinfo{publisher}{Physics of Magnetic Semiconductors (MIR Publishers,
  Moscow)}, \bibinfo{year}{1983}).

\bibitem[{\citenamefont{Furdyna}(1988)}]{furdyna1988:JAP}
\bibinfo{author}{\bibfnamefont{J.~K.} \bibnamefont{Furdyna}},
  \bibinfo{journal}{J. Appl. Phys.} \textbf{\bibinfo{volume}{64}},
  \bibinfo{pages}{R29} (\bibinfo{year}{1988}).

\bibitem[{\citenamefont{Amasha et~al.}(2008)\citenamefont{Amasha, MacLean,
  Radu, {Zumb\"uhl}, Kastner, Hanson, and Gossard}}]{amasha2008:PRL}
\bibinfo{author}{\bibfnamefont{S.}~\bibnamefont{Amasha}},
  \bibinfo{author}{\bibfnamefont{K.}~\bibnamefont{MacLean}},
  \bibinfo{author}{\bibfnamefont{I.~P.} \bibnamefont{Radu}},
  \bibinfo{author}{\bibfnamefont{D.~M.} \bibnamefont{{Zumb\"uhl}}},
  \bibinfo{author}{\bibfnamefont{M.~A.} \bibnamefont{Kastner}},
  \bibinfo{author}{\bibfnamefont{M.~P.} \bibnamefont{Hanson}},
  \bibnamefont{and} \bibinfo{author}{\bibfnamefont{A.~C.}
  \bibnamefont{Gossard}}, \bibinfo{journal}{Phys. Rev. Lett.}
  \textbf{\bibinfo{volume}{100}}, \bibinfo{pages}{046803}
  (\bibinfo{year}{2008}).

\bibitem[{\citenamefont{Paget et~al.}(1977)\citenamefont{Paget, Lampel,
  Sapoval, and Safarov}}]{paget1977:PRB}
\bibinfo{author}{\bibfnamefont{D.}~\bibnamefont{Paget}},
  \bibinfo{author}{\bibfnamefont{G.}~\bibnamefont{Lampel}},
  \bibinfo{author}{\bibfnamefont{B.}~\bibnamefont{Sapoval}}, \bibnamefont{and}
  \bibinfo{author}{\bibfnamefont{V.~I.} \bibnamefont{Safarov}},
  \bibinfo{journal}{Phys. Rev. B} \textbf{\bibinfo{volume}{15}},
  \bibinfo{pages}{5780} (\bibinfo{year}{1977}).

\bibitem[{\citenamefont{Kuo et~al.}(2006)\citenamefont{Kuo, Hsu, Shen, Chiu,
  Fan, Lin, Chia, Chou, Yasar, Mallory et~al.}}]{kuo2006:APL}
\bibinfo{author}{\bibfnamefont{M.~C.} \bibnamefont{Kuo}},
  \bibinfo{author}{\bibfnamefont{J.~S.} \bibnamefont{Hsu}},
  \bibinfo{author}{\bibfnamefont{J.~L.} \bibnamefont{Shen}},
  \bibinfo{author}{\bibfnamefont{K.~C.} \bibnamefont{Chiu}},
  \bibinfo{author}{\bibfnamefont{W.~C.} \bibnamefont{Fan}},
  \bibinfo{author}{\bibfnamefont{Y.~C.} \bibnamefont{Lin}},
  \bibinfo{author}{\bibfnamefont{C.~H.} \bibnamefont{Chia}},
  \bibinfo{author}{\bibfnamefont{W.~C.} \bibnamefont{Chou}},
  \bibinfo{author}{\bibfnamefont{M.}~\bibnamefont{Yasar}},
  \bibinfo{author}{\bibfnamefont{R.}~\bibnamefont{Mallory}},
  \bibnamefont{et~al.}, \bibinfo{journal}{Appl. Phys. Lett.}
  \textbf{\bibinfo{volume}{89}}, \bibinfo{pages}{263111}
  (\bibinfo{year}{2006}).

\bibitem[{foo()}]{footnote7}
\bibinfo{note}{This is further corroborated by applying ZnTe paramaters to the
  theoretical approach from Ref.~\onlinecite{vyborny2012:PRB}}.

\bibitem[{\citenamefont{deSousa}(2009)}]{desousa2009:TAP}
\bibinfo{author}{\bibfnamefont{R.}~\bibnamefont{deSousa}},
  \bibinfo{journal}{in: M. Fanciulli (Ed.), Topics Appl. Physics}
  \textbf{\bibinfo{volume}{115}}, \bibinfo{pages}{183} (\bibinfo{year}{2009}).

\bibitem[{\citenamefont{Dietl et~al.}(1995)\citenamefont{Dietl, Peyla,
  Grieshaber, and d\'Aubign}}]{dietl1995:JMMM}
\bibinfo{author}{\bibfnamefont{T.}~\bibnamefont{Dietl}},
  \bibinfo{author}{\bibfnamefont{P.}~\bibnamefont{Peyla}},
  \bibinfo{author}{\bibfnamefont{W.}~\bibnamefont{Grieshaber}},
  \bibnamefont{and}
  \bibinfo{author}{\bibfnamefont{M.}~\bibnamefont{d\'Aubign}},
  \bibinfo{journal}{J. Magn. Magn. Mater.} \textbf{\bibinfo{volume}{140-144}},
  \bibinfo{pages}{2051} (\bibinfo{year}{1995}).

\bibitem[{\citenamefont{Elzerman et~al.}(2004)\citenamefont{Elzerman, Hanson,
  {van Beveren}, Witkamp, Vandersypen, and Kouwenhoven}}]{elzerman2004:N}
\bibinfo{author}{\bibfnamefont{J.~M.} \bibnamefont{Elzerman}},
  \bibinfo{author}{\bibfnamefont{R.}~\bibnamefont{Hanson}},
  \bibinfo{author}{\bibfnamefont{L.~H.~W.} \bibnamefont{{van Beveren}}},
  \bibinfo{author}{\bibfnamefont{B.}~\bibnamefont{Witkamp}},
  \bibinfo{author}{\bibfnamefont{L.~M.~K.} \bibnamefont{Vandersypen}},
  \bibnamefont{and} \bibinfo{author}{\bibfnamefont{L.~P.}
  \bibnamefont{Kouwenhoven}}, \bibinfo{journal}{Nature}
  \textbf{\bibinfo{volume}{430}}, \bibinfo{pages}{431} (\bibinfo{year}{2004}).

\bibitem[{\citenamefont{Aleiner and Fa\v{l}ko}(2001)}]{aleiner2001:PRL}
\bibinfo{author}{\bibfnamefont{I.~L.} \bibnamefont{Aleiner}} \bibnamefont{and}
  \bibinfo{author}{\bibfnamefont{V.~I.} \bibnamefont{Fa\v{l}ko}},
  \bibinfo{journal}{Phys. Rev. Lett.} \textbf{\bibinfo{volume}{87}},
  \bibinfo{pages}{256801} (\bibinfo{year}{2001}).

\bibitem[{\citenamefont{Levitov and Rashba}(2003)}]{levitov2003:PRB}
\bibinfo{author}{\bibfnamefont{L.~S.} \bibnamefont{Levitov}} \bibnamefont{and}
  \bibinfo{author}{\bibfnamefont{E.~I.} \bibnamefont{Rashba}},
  \bibinfo{journal}{Phys. Rev. B} \textbf{\bibinfo{volume}{67}},
  \bibinfo{pages}{115324} (\bibinfo{year}{2003}).

\bibitem[{\citenamefont{Stepanenko et~al.}(2003)\citenamefont{Stepanenko,
  Bonesteel, DiVincenzo, Burkard, and Loss}}]{stepanenko2003:PRB}
\bibinfo{author}{\bibfnamefont{D.}~\bibnamefont{Stepanenko}},
  \bibinfo{author}{\bibfnamefont{N.~E.} \bibnamefont{Bonesteel}},
  \bibinfo{author}{\bibfnamefont{D.~P.} \bibnamefont{DiVincenzo}},
  \bibinfo{author}{\bibfnamefont{G.}~\bibnamefont{Burkard}}, \bibnamefont{and}
  \bibinfo{author}{\bibfnamefont{D.}~\bibnamefont{Loss}},
  \bibinfo{journal}{Phys. Rev. B} \textbf{\bibinfo{volume}{68}},
  \bibinfo{pages}{115306} (\bibinfo{year}{2003}).

\bibitem[{\citenamefont{Schliemann
  et~al.}(2003{\natexlab{b}})\citenamefont{Schliemann, Egues, and
  Loss}}]{schliemann2003:PRL}
\bibinfo{author}{\bibfnamefont{J.}~\bibnamefont{Schliemann}},
  \bibinfo{author}{\bibfnamefont{J.~C.} \bibnamefont{Egues}}, \bibnamefont{and}
  \bibinfo{author}{\bibfnamefont{D.}~\bibnamefont{Loss}},
  \bibinfo{journal}{Phys. Rev. Lett.} \textbf{\bibinfo{volume}{90}},
  \bibinfo{pages}{146801} (\bibinfo{year}{2003}{\natexlab{b}}).

\bibitem[{\citenamefont{Tokatly and
  Sherman}(2010{\natexlab{a}})}]{tokatly2010:AP}
\bibinfo{author}{\bibfnamefont{I.~V.} \bibnamefont{Tokatly}} \bibnamefont{and}
  \bibinfo{author}{\bibfnamefont{E.~Y.} \bibnamefont{Sherman}},
  \bibinfo{journal}{Annals of Physics} \textbf{\bibinfo{volume}{325}},
  \bibinfo{pages}{1104} (\bibinfo{year}{2010}{\natexlab{a}}).

\bibitem[{\citenamefont{Tokatly and
  Sherman}(2010{\natexlab{b}})}]{tokatly2010:PRB}
\bibinfo{author}{\bibfnamefont{I.~V.} \bibnamefont{Tokatly}} \bibnamefont{and}
  \bibinfo{author}{\bibfnamefont{E.~Y.} \bibnamefont{Sherman}},
  \bibinfo{journal}{Phys. Rev. B} \textbf{\bibinfo{volume}{82}},
  \bibinfo{pages}{161305} (\bibinfo{year}{2010}{\natexlab{b}}).

\bibitem[{\citenamefont{Stano and Fabian}(2006)}]{stano2006:PRL}
\bibinfo{author}{\bibfnamefont{P.}~\bibnamefont{Stano}} \bibnamefont{and}
  \bibinfo{author}{\bibfnamefont{J.}~\bibnamefont{Fabian}},
  \bibinfo{journal}{Phys. Rev. Lett.} \textbf{\bibinfo{volume}{96}},
  \bibinfo{pages}{186602} (\bibinfo{year}{2006}).

\bibitem[{\citenamefont{Baruffa et~al.}(2010)\citenamefont{Baruffa, Stano, and
  Fabian}}]{baruffa2010:PRL}
\bibinfo{author}{\bibfnamefont{F.}~\bibnamefont{Baruffa}},
  \bibinfo{author}{\bibfnamefont{P.}~\bibnamefont{Stano}}, \bibnamefont{and}
  \bibinfo{author}{\bibfnamefont{J.}~\bibnamefont{Fabian}},
  \bibinfo{journal}{Phys. Rev. Lett.} \textbf{\bibinfo{volume}{104}},
  \bibinfo{pages}{126401} (\bibinfo{year}{2010}).

\bibitem[{\citenamefont{Oswaldowski
  et~al.}(unpublished)\citenamefont{Oswaldowski, Stano, Pletyukov, and
  {\v{Z}uti\'c}}}]{oswaldowski2012:U}
\bibinfo{author}{\bibfnamefont{R.}~\bibnamefont{Oswaldowski}},
  \bibinfo{author}{\bibfnamefont{P.}~\bibnamefont{Stano}},
  \bibinfo{author}{\bibfnamefont{A.}~\bibnamefont{Pletyukov}},
  \bibnamefont{and}
  \bibinfo{author}{\bibfnamefont{I.}~\bibnamefont{{\v{Z}uti\'c}}}
  (\bibinfo{year}{unpublished}).

\bibitem[{\citenamefont{Fischer et~al.}(2008)\citenamefont{Fischer, Coish,
  Bulaev, and Loss}}]{fischer2008:PRB}
\bibinfo{author}{\bibfnamefont{J.}~\bibnamefont{Fischer}},
  \bibinfo{author}{\bibfnamefont{W.~A.} \bibnamefont{Coish}},
  \bibinfo{author}{\bibfnamefont{D.~V.} \bibnamefont{Bulaev}},
  \bibnamefont{and} \bibinfo{author}{\bibfnamefont{D.}~\bibnamefont{Loss}},
  \bibinfo{journal}{Phys. Rev. B} \textbf{\bibinfo{volume}{78}},
  \bibinfo{pages}{155329} (\bibinfo{year}{2008}).

\bibitem[{\citenamefont{Pfund et~al.}(2007)\citenamefont{Pfund, Shorubalko,
  Ensslin, and Leturcq}}]{pfund2007:PRL}
\bibinfo{author}{\bibfnamefont{A.}~\bibnamefont{Pfund}},
  \bibinfo{author}{\bibfnamefont{I.}~\bibnamefont{Shorubalko}},
  \bibinfo{author}{\bibfnamefont{K.}~\bibnamefont{Ensslin}}, \bibnamefont{and}
  \bibinfo{author}{\bibfnamefont{R.}~\bibnamefont{Leturcq}},
  \bibinfo{journal}{Phys. Rev. Lett.} \textbf{\bibinfo{volume}{99}},
  \bibinfo{pages}{036801} (\bibinfo{year}{2007}).

\bibitem[{\citenamefont{Rudner and Levitov}(2007)}]{rudner2007:PRL}
\bibinfo{author}{\bibfnamefont{M.~S.} \bibnamefont{Rudner}} \bibnamefont{and}
  \bibinfo{author}{\bibfnamefont{L.~S.} \bibnamefont{Levitov}},
  \bibinfo{journal}{Phys. Rev. Lett.} \textbf{\bibinfo{volume}{99}},
  \bibinfo{pages}{036602} (\bibinfo{year}{2007}).

\bibitem[{\citenamefont{Erlingsson and Nazarov}(2002)}]{erlingsson2002:PRB}
\bibinfo{author}{\bibfnamefont{S.~I.} \bibnamefont{Erlingsson}}
  \bibnamefont{and} \bibinfo{author}{\bibfnamefont{Y.~V.}
  \bibnamefont{Nazarov}}, \bibinfo{journal}{Phys. Rev. B}
  \textbf{\bibinfo{volume}{66}}, \bibinfo{pages}{155327}
  (\bibinfo{year}{2002}).

\bibitem[{\citenamefont{Shenvi et~al.}(2005)\citenamefont{Shenvi, {de Sousa},
  and Whaley}}]{shenvi2005:PRB}
\bibinfo{author}{\bibfnamefont{N.}~\bibnamefont{Shenvi}},
  \bibinfo{author}{\bibfnamefont{R.}~\bibnamefont{{de Sousa}}},
  \bibnamefont{and} \bibinfo{author}{\bibfnamefont{K.~B.}
  \bibnamefont{Whaley}}, \bibinfo{journal}{Phys. Rev. B}
  \textbf{\bibinfo{volume}{71}}, \bibinfo{pages}{224411}
  (\bibinfo{year}{2005}).

\bibitem[{\citenamefont{Yao et~al.}(2006)\citenamefont{Yao, Liu, and
  Sham}}]{yao2006:PRB}
\bibinfo{author}{\bibfnamefont{W.}~\bibnamefont{Yao}},
  \bibinfo{author}{\bibfnamefont{R.-B.} \bibnamefont{Liu}}, \bibnamefont{and}
  \bibinfo{author}{\bibfnamefont{L.~J.} \bibnamefont{Sham}},
  \bibinfo{journal}{Phys. Rev. B} \textbf{\bibinfo{volume}{74}},
  \bibinfo{pages}{195301} (\bibinfo{year}{2006}).

\bibitem[{\citenamefont{{L\"owdin}}(1951)}]{lowdin1951:JCP}
\bibinfo{author}{\bibfnamefont{P.}~\bibnamefont{{L\"owdin}}},
  \bibinfo{journal}{J. Chem. Phys.} \textbf{\bibinfo{volume}{19}},
  \bibinfo{pages}{1396} (\bibinfo{year}{1951}).

\bibitem[{\citenamefont{Bir and Pikus}(1974)}]{bir-pikus}
\bibinfo{author}{\bibfnamefont{G.~L.} \bibnamefont{Bir}} \bibnamefont{and}
  \bibinfo{author}{\bibfnamefont{G.~E.} \bibnamefont{Pikus}},
  \emph{\bibinfo{title}{Symmetry and Strain-induced Effects in Semiconductors}}
  (\bibinfo{publisher}{Wiley/Halsted Press}, \bibinfo{year}{1974}).

\bibitem[{\citenamefont{Abragam}(1961)}]{abragam}
\bibinfo{author}{\bibfnamefont{A.}~\bibnamefont{Abragam}},
  \emph{\bibinfo{title}{The Principles of Nuclear Magnetism}}
  (\bibinfo{publisher}{Oxford University Press}, \bibinfo{year}{1961}).

\bibitem[{\citenamefont{Abolfath et~al.}(2008)\citenamefont{Abolfath, Petukhov,
  and {\v{Z}uti\'c}}}]{abolfath2008:PRL}
\bibinfo{author}{\bibfnamefont{R.~M.} \bibnamefont{Abolfath}},
  \bibinfo{author}{\bibfnamefont{A.~G.} \bibnamefont{Petukhov}},
  \bibnamefont{and}
  \bibinfo{author}{\bibfnamefont{I.}~\bibnamefont{{\v{Z}uti\'c}}},
  \bibinfo{journal}{Phys. Rev. Lett.} \textbf{\bibinfo{volume}{101}},
  \bibinfo{pages}{207202} (\bibinfo{year}{2008}).

\bibitem[{\citenamefont{Austing et~al.}(1999)\citenamefont{Austing, Sasaki,
  Tarucha, Reimann, Koskinen, and Manninen}}]{austing1999:PRB}
\bibinfo{author}{\bibfnamefont{D.~G.} \bibnamefont{Austing}},
  \bibinfo{author}{\bibfnamefont{S.}~\bibnamefont{Sasaki}},
  \bibinfo{author}{\bibfnamefont{S.}~\bibnamefont{Tarucha}},
  \bibinfo{author}{\bibfnamefont{S.~M.} \bibnamefont{Reimann}},
  \bibinfo{author}{\bibfnamefont{M.}~\bibnamefont{Koskinen}}, \bibnamefont{and}
  \bibinfo{author}{\bibfnamefont{M.}~\bibnamefont{Manninen}},
  \bibinfo{journal}{Phys. Rev. B} \textbf{\bibinfo{volume}{60}},
  \bibinfo{pages}{11514} (\bibinfo{year}{1999}).

\bibitem[{\citenamefont{Henneberger and Puls}(2010)}]{henneberger}
\bibinfo{author}{\bibfnamefont{F.}~\bibnamefont{Henneberger}} \bibnamefont{and}
  \bibinfo{author}{\bibfnamefont{J.}~\bibnamefont{Puls}},
  \emph{\bibinfo{title}{Diluted Magnetic Quantum Dots}}
  (\bibinfo{publisher}{Introduction to the Physics of Diluted Magnetic
  Semiconductors edited by J. Kossut and J. A. Gaj, Springer, Berlin},
  \bibinfo{year}{2010}).

\bibitem[{\citenamefont{Yakovlev and Merkulov}(2010)}]{yakovlev2}
\bibinfo{author}{\bibfnamefont{D.~R.} \bibnamefont{Yakovlev}} \bibnamefont{and}
  \bibinfo{author}{\bibfnamefont{I.~A.} \bibnamefont{Merkulov}},
  \emph{\bibinfo{title}{Spin and Energy Transfer Between Carriers, Magnetic
  Ions, and Lattice}} (\bibinfo{publisher}{Introduction to the Physics of
  Diluted Magnetic Semiconductors edited by J. Kossut and J. A. Gaj, Springer,
  Berlin}, \bibinfo{year}{2010}).

\bibitem[{\citenamefont{Klimov}(2007)}]{klimov2007:ARPCH}
\bibinfo{author}{\bibfnamefont{V.~I.} \bibnamefont{Klimov}},
  \bibinfo{journal}{Annu. Rev. Phys. Chem.} \textbf{\bibinfo{volume}{58}},
  \bibinfo{pages}{635} (\bibinfo{year}{2007}).

\bibitem[{\citenamefont{Scholes}(2007)}]{scholes2008:AFM}
\bibinfo{author}{\bibfnamefont{G.~D.} \bibnamefont{Scholes}},
  \bibinfo{journal}{Adv. Funct. Mater.} \textbf{\bibinfo{volume}{18}},
  \bibinfo{pages}{1157} (\bibinfo{year}{2007}).

\bibitem[{\citenamefont{Stern et~al.}(2005)\citenamefont{Stern, Poggio, Bartl,
  Hu, Stucky, and Awschalom}}]{stern2005:PRB}
\bibinfo{author}{\bibfnamefont{N.~P.} \bibnamefont{Stern}},
  \bibinfo{author}{\bibfnamefont{M.}~\bibnamefont{Poggio}},
  \bibinfo{author}{\bibfnamefont{M.~H.} \bibnamefont{Bartl}},
  \bibinfo{author}{\bibfnamefont{E.~L.} \bibnamefont{Hu}},
  \bibinfo{author}{\bibfnamefont{G.~D.} \bibnamefont{Stucky}},
  \bibnamefont{and} \bibinfo{author}{\bibfnamefont{D.~D.}
  \bibnamefont{Awschalom}}, \bibinfo{journal}{Phys. Rev. B}
  \textbf{\bibinfo{volume}{72}}, \bibinfo{pages}{161303}
  (\bibinfo{year}{2005}).

\bibitem[{\citenamefont{Beaulac et~al.}(2008)\citenamefont{Beaulac, Archer,
  Ochsenbein, and Gamelin}}]{beaulac2008:AFM}
\bibinfo{author}{\bibfnamefont{R.}~\bibnamefont{Beaulac}},
  \bibinfo{author}{\bibfnamefont{P.~I.} \bibnamefont{Archer}},
  \bibinfo{author}{\bibfnamefont{S.~T.} \bibnamefont{Ochsenbein}},
  \bibnamefont{and} \bibinfo{author}{\bibfnamefont{D.~R.}
  \bibnamefont{Gamelin}}, \bibinfo{journal}{Adv. Funct. Mater.}
  \textbf{\bibinfo{volume}{18}}, \bibinfo{pages}{3873} (\bibinfo{year}{2008}).

\bibitem[{\citenamefont{Bussian et~al.}(2009)\citenamefont{Bussian, Crooker,
  Yin, Brynda, Efros, and Klimov}}]{bussian2009:NM}
\bibinfo{author}{\bibfnamefont{D.~A.} \bibnamefont{Bussian}},
  \bibinfo{author}{\bibfnamefont{S.~A.} \bibnamefont{Crooker}},
  \bibinfo{author}{\bibfnamefont{M.}~\bibnamefont{Yin}},
  \bibinfo{author}{\bibfnamefont{M.}~\bibnamefont{Brynda}},
  \bibinfo{author}{\bibfnamefont{A.~L.} \bibnamefont{Efros}}, \bibnamefont{and}
  \bibinfo{author}{\bibfnamefont{V.~I.} \bibnamefont{Klimov}},
  \bibinfo{journal}{Nature Mater.} \textbf{\bibinfo{volume}{8}},
  \bibinfo{pages}{35} (\bibinfo{year}{2009}).

\bibitem[{\citenamefont{Ochsenbein et~al.}(2009)\citenamefont{Ochsenbein, Feng,
  Whitaker, Badaeva, Liu, Li, and Gamelin}}]{ochsenbein2009:NN}
\bibinfo{author}{\bibfnamefont{S.~T.} \bibnamefont{Ochsenbein}},
  \bibinfo{author}{\bibfnamefont{Y.}~\bibnamefont{Feng}},
  \bibinfo{author}{\bibfnamefont{K.~M.} \bibnamefont{Whitaker}},
  \bibinfo{author}{\bibfnamefont{E.}~\bibnamefont{Badaeva}},
  \bibinfo{author}{\bibfnamefont{W.~K.} \bibnamefont{Liu}},
  \bibinfo{author}{\bibfnamefont{X.}~\bibnamefont{Li}}, \bibnamefont{and}
  \bibinfo{author}{\bibfnamefont{D.~R.} \bibnamefont{Gamelin}},
  \bibinfo{journal}{Nature Nanotech.} \textbf{\bibinfo{volume}{4}},
  \bibinfo{pages}{681} (\bibinfo{year}{2009}).

\bibitem[{\citenamefont{{\v{Z}uti\'c} and Petukhov}(2009)}]{zutic2009:NN}
\bibinfo{author}{\bibfnamefont{I.}~\bibnamefont{{\v{Z}uti\'c}}}
  \bibnamefont{and} \bibinfo{author}{\bibfnamefont{A.~G.}
  \bibnamefont{Petukhov}}, \bibinfo{journal}{Nature Nanotech.}
  \textbf{\bibinfo{volume}{4}}, \bibinfo{pages}{623} (\bibinfo{year}{2009}).

\bibitem[{\citenamefont{Viswanatha et~al.}(2011)\citenamefont{Viswanatha,
  Pietryga, Klimov, and Crooker}}]{viswanatha2011:PRL}
\bibinfo{author}{\bibfnamefont{R.}~\bibnamefont{Viswanatha}},
  \bibinfo{author}{\bibfnamefont{J.~M.} \bibnamefont{Pietryga}},
  \bibinfo{author}{\bibfnamefont{V.~I.} \bibnamefont{Klimov}},
  \bibnamefont{and} \bibinfo{author}{\bibfnamefont{S.~A.}
  \bibnamefont{Crooker}}, \bibinfo{journal}{Phys. Rev. Lett.}
  \textbf{\bibinfo{volume}{107}}, \bibinfo{pages}{067402}
  (\bibinfo{year}{2011}).

\bibitem[{\citenamefont{Winkler}(2003)}]{winkler}
\bibinfo{author}{\bibfnamefont{R.}~\bibnamefont{Winkler}},
  \emph{\bibinfo{title}{Spin–Orbit Coupling Effects in Two-Dimensional
  Electron and Hole Systems}} (\bibinfo{publisher}{Springer-Verlag},
  \bibinfo{year}{2003}).

\bibitem[{\citenamefont{Bhattacharjee}(2007)}]{bhattacharjee2007:PRB}
\bibinfo{author}{\bibfnamefont{A.~K.} \bibnamefont{Bhattacharjee}},
  \bibinfo{journal}{Phys. Rev. B} \textbf{\bibinfo{volume}{76}},
  \bibinfo{pages}{075305} (\bibinfo{year}{2007}).

\bibitem[{\citenamefont{Fischer and Loss}(2010)}]{fischer2010:PRL}
\bibinfo{author}{\bibfnamefont{J.}~\bibnamefont{Fischer}} \bibnamefont{and}
  \bibinfo{author}{\bibfnamefont{D.}~\bibnamefont{Loss}},
  \bibinfo{journal}{Phys. Rev. Lett.} \textbf{\bibinfo{volume}{105}},
  \bibinfo{pages}{266603} (\bibinfo{year}{2010}).

\bibitem[{\citenamefont{Katsaros et~al.}(unpublished)\citenamefont{Katsaros,
  Golovach, Spathis, Ares, Stoffel, Fournel, Schmidt, Glazman, and {De
  Franceschi}}}]{katsaros2011:CM}
\bibinfo{author}{\bibfnamefont{G.}~\bibnamefont{Katsaros}},
  \bibinfo{author}{\bibfnamefont{V.~N.} \bibnamefont{Golovach}},
  \bibinfo{author}{\bibfnamefont{P.}~\bibnamefont{Spathis}},
  \bibinfo{author}{\bibfnamefont{N.}~\bibnamefont{Ares}},
  \bibinfo{author}{\bibfnamefont{M.}~\bibnamefont{Stoffel}},
  \bibinfo{author}{\bibfnamefont{F.}~\bibnamefont{Fournel}},
  \bibinfo{author}{\bibfnamefont{O.~G.} \bibnamefont{Schmidt}},
  \bibinfo{author}{\bibfnamefont{L.~I.} \bibnamefont{Glazman}},
  \bibnamefont{and} \bibinfo{author}{\bibfnamefont{S.}~\bibnamefont{{De
  Franceschi}}}, \bibinfo{journal}{arxiv:1107.3919}
  (\bibinfo{year}{unpublished}).

\bibitem[{\citenamefont{Binggeli and Baldereschi}(1991)}]{binggeli1991:PRB}
\bibinfo{author}{\bibfnamefont{N.}~\bibnamefont{Binggeli}} \bibnamefont{and}
  \bibinfo{author}{\bibfnamefont{A.}~\bibnamefont{Baldereschi}},
  \bibinfo{journal}{Phys. Rev. B} \textbf{\bibinfo{volume}{43}},
  \bibinfo{pages}{14734} (\bibinfo{year}{1991}).

\bibitem[{\citenamefont{Shanabrook et~al.}(1989)\citenamefont{Shanabrook,
  Glembocki, Broido, and Wang}}]{shanabrook1989:PRB}
\bibinfo{author}{\bibfnamefont{B.~V.} \bibnamefont{Shanabrook}},
  \bibinfo{author}{\bibfnamefont{O.~J.} \bibnamefont{Glembocki}},
  \bibinfo{author}{\bibfnamefont{D.~A.} \bibnamefont{Broido}},
  \bibnamefont{and} \bibinfo{author}{\bibfnamefont{W.~I.} \bibnamefont{Wang}},
  \bibinfo{journal}{Phys. Rev. B} \textbf{\bibinfo{volume}{39}},
  \bibinfo{pages}{3411} (\bibinfo{year}{1989}).

\bibitem[{\citenamefont{Friedrich et~al.}(1994)\citenamefont{Friedrich, Kraus,
  Schaack, and Schmitt}}]{friedrich1994:JPCM}
\bibinfo{author}{\bibfnamefont{T.}~\bibnamefont{Friedrich}},
  \bibinfo{author}{\bibfnamefont{J.}~\bibnamefont{Kraus}},
  \bibinfo{author}{\bibfnamefont{G.}~\bibnamefont{Schaack}}, \bibnamefont{and}
  \bibinfo{author}{\bibfnamefont{W.~O.~G.} \bibnamefont{Schmitt}},
  \bibinfo{journal}{J. Phys.: Condens. Matter} \textbf{\bibinfo{volume}{6}},
  \bibinfo{pages}{4307} (\bibinfo{year}{1994}).

\bibitem[{\citenamefont{Wagner et~al.}(1992)\citenamefont{Wagner, Lankes, Wolf,
  Kuhn, Link, and Gebhardt}}]{wagner1992:JCG}
\bibinfo{author}{\bibfnamefont{H.}~\bibnamefont{Wagner}},
  \bibinfo{author}{\bibfnamefont{S.}~\bibnamefont{Lankes}},
  \bibinfo{author}{\bibfnamefont{K.}~\bibnamefont{Wolf}},
  \bibinfo{author}{\bibfnamefont{W.}~\bibnamefont{Kuhn}},
  \bibinfo{author}{\bibfnamefont{P.}~\bibnamefont{Link}}, \bibnamefont{and}
  \bibinfo{author}{\bibfnamefont{W.}~\bibnamefont{Gebhardt}},
  \bibinfo{journal}{J. Cryst. Growth} \textbf{\bibinfo{volume}{117}},
  \bibinfo{pages}{303} (\bibinfo{year}{1992}).

\bibitem[{\citenamefont{Vyborny et~al.}(2012)\citenamefont{Vyborny, Han,
  Oszwa{\l}dowski, \v{Z}uti{\'c}, and Petukhov}}]{vyborny2012:PRB}
\bibinfo{author}{\bibfnamefont{K.}~\bibnamefont{Vyborny}},
  \bibinfo{author}{\bibfnamefont{J.~E.} \bibnamefont{Han}},
  \bibinfo{author}{\bibfnamefont{R.}~\bibnamefont{Oszwa{\l}dowski}},
  \bibinfo{author}{\bibfnamefont{I.}~\bibnamefont{\v{Z}uti{\'c}}},
  \bibnamefont{and} \bibinfo{author}{\bibfnamefont{A.~G.}
  \bibnamefont{Petukhov}}, \bibinfo{journal}{Phys. Rev. B}
  \textbf{\bibinfo{volume}{85}}, \bibinfo{pages}{155312}
  (\bibinfo{year}{2012}).

\end{thebibliography}
\bibliographystyle{apsrev}

\end{document}